\documentclass[12pt]{article}
\usepackage[margin=1in]{geometry}
\usepackage[small]{titlesec}
\usepackage{natbib,amsmath,amssymb,amsthm,amstext,graphicx,setspace,paralist,booktabs,rotating,subcaption,times}
\usepackage{algorithm, algorithmic, comment}
\usepackage{color,xcolor,bm,enumerate}

\usepackage{pb-diagram}
\usepackage{pifont}
\usepackage{appendix}
\usepackage{indentfirst,multirow,hhline,tabularx,array}
\usepackage{amsfonts,amsopn,mathrsfs}
\usepackage{upgreek,url}

\AtBeginDocument{%
	\abovedisplayskip=5pt plus 2pt minus 2pt
	\belowdisplayskip=\abovedisplayskip
	\abovedisplayshortskip=2pt plus 2pt minus 2pt
	\belowdisplayshortskip=\belowdisplayskip}

\titlespacing*{\section}{0pt}{*2}{*1}
\titlespacing*{\subsection}{0pt}{*2}{*1}

\bibpunct{(}{)}{;}{a}{}{,}
\allowdisplaybreaks[1]

\newtheorem{theorem}{Theorem}
\newtheorem{lemma}{Lemma}

\theoremstyle{definition}

\def\R{{\mathbb R}}

\def\H{{\mathcal H}}
\def\E{{\mathbb E}}

\def\cov{{\rm Cov}}
\def\var{{\rm Var}}
\def\supp{{\rm supp}}

\def\H{{\cal{H}}}

\newcommand{\bX}{\mathbf{X}}

\newcommand{\bZ}{\mathbf{Z}}
\newcommand{\bE}{\mathbf{E}}
\newcommand{\bD}{\mathbf{D}}

\newcommand{\bM}{\mathbf{M}}

\newcommand{\wbX}{\widehat{\mathbf{X}}}
\newcommand{\wbD}{\widehat{\mathbf{D}}}
\newcommand{\wbbeta}{\widehat{\bm{\beta}}}

\newcommand{\wbtheta}{\widehat{\bm{\theta}}}
\newcommand{\wbGamma}{\widehat{\bm{\Gamma}}}

\newcommand{\bbeta}{\bm{\beta}}
\newcommand{\bmeta}{\bm{\eta}}
\newcommand{\bzeta}{\bm{\zeta}}
\newcommand{\btheta}{\bm{\theta}}
\newcommand{\bdelta}{\bm{\delta}}
\newcommand{\bGamma}{\bm{\Gamma}}
\newcommand{\bgamma}{\bm{\gamma}}
\newcommand{\bSigma}{\bm{\Sigma}}
\newcommand{\bOmega}{\bm{\Omega}}

\DeclareMathOperator*{\argmin}{argmin}

\begin{document}
	
\begin{titlepage}
\setstretch{1.24}
\title{Hypothesis Testing in High-Dimensional Instrumental Variables Regression with an Application to Genomics Data}
\author{
	Jiarui Lu$^{1*}$,
	Hongzhe Li$^{1}$
}
\date{}
\maketitle
\thispagestyle{empty}

\footnotetext[1]{Department of Biostatistics, Epidemiology and Informatics, Perelman School of Medicine, University of Pennsylvania, Philadelphia, PA 19104. }
\footnotetext[2]{Email: jiaruilu@pennmedicine.upenn.edu}

\begin{abstract}
Gene expression and phenotype association can be affected by potential unmeasured confounders from multiple sources,  leading to biased estimates of the associations. Since genetic variants largely explain gene expression variations, they can be used as instruments  in studying the association between gene expressions and phenotype in the framework of high dimensional instrumental variable (IV) regression. However,  because the dimensions of  both genetic variants and gene expressions  are often larger  than the sample size,  statistical inferences such as hypothesis testing for  such high dimensional IV models are not trivial and have not been investigated in literature. The problem is more challenging since  the instrumental variables (e.g., genetic variants) have to be selected among a large set of genetic variants. This paper considers the problem of hypothesis testing for sparse IV regression models and presents methods for testing single regression coefficient and multiple testing of multiple coefficients, where  the test statistic for each single coefficient is constructed based on an inverse regression. A multiple testing procedure is developed for selecting variables and is  shown to  control the false discovery rate. Simulations are conducted to evaluate the performance of our proposed methods. These methods are illustrated by an analysis of a yeast dataset in order to   identify genes that are associated with growth in the presence of hydrogen peroxide.

\end{abstract}

\end{titlepage}

\section{Introduction}
Many genomic studies collect both germline genetic variants and tissue-specific gene expression data on the same set of individuals  in order to understand how genetic variants perturb gene expressions that lead to clinical phenotypes. Among various methods, association analysis between gene expression and phenotype such as differential gene expression analysis has been widely reported. Such  studies have shown that gene expressions are associated  with many common human diseases, such as liver disease  \citep{romeo2008genetic, speliotes2011genome} and heart failure \citep{HFexp}. However,  there are possibly  many  unmeasured factors that affect both gene expressions and  phenotype of interest \citep{leek2007capturing,hoggart2003control}. The existence of such unmeasured confounding variables  can cause correlation between the error term and one or some of the independent variables and lead to identifying false associations. Particularly, the independence assumption between gene expressions  and errors are required in linear regression in order to obtain valid statistical inference of  the effects of gene expressions on phenotype. If this assumption is violated, standard methods can lead to biased estimates \citep{lin2015regularization, fan2014endogeneity}.

One way to deal with unmeasured confounding is to apply  instrumental variables (IV) regression, which has been studied  extensively in low dimensional settings \citep{imbens2014instrumental}.  In the context of our applications, we  treat genetic variants as instrumental variables  in studying the association between gene expressions and phenotypes.  Standard method to fit the IV models is to apply  two-stage regressions  to obtain valid estimation of the true parameters.  However, in genetical genomics studies,  the dimensions of both genetic variants and gene expressions are much larger than the sample sizes,  making the classic two-stage regression methods of fitting the IV models infeasible. To account for high dimensionality, penalized regression methods have been developed  to select the  instruments in the first stage and then to  select gene expressions in the second stage \citep{lin2015regularization}.  \cite{lin2015regularization}  provided the estimation error bounds of proposed two-stage estimators and but did not study the related  problem of statistical inference.

For linear regression models  in high-dimensional setting,   \cite{javanmard2014confidence} developed  a de-biased procedure  to construct an asymptotically normally distributed estimator  based on the original  biased Lasso estimator. The asymptotic results can be used for hypothesis testing. \citet{zhang2014confidence} proposed a low-dimensional projection estimator to correct the bias, sharing  a similar idea as  \citet{javanmard2014confidence}. In a more general framework, \citet{ning2017} considered the hypothesis testing problem for general penalized M-estimator, where they constructed a decorrelated score statistic in  high-dimensional setting. All these methods for high dimensional  linear regression inference require the critical assumption that the error terms are independent of  the covariates, and therefore cannot  be  applied to the IV models directly.

This paper presents  methods for hypothesis testing for high dimensional IV models, including statistical test of a single regression coefficient and  a multiple testing procedure for variable selection. The methods build on the work of \citet{lin2015regularization} to obtain a consistent estimator of the regression coefficients, and  the work of \citet{liu2013gaussian} to perform inverse regressions to construct the bias-corrected test statistics. The idea of inverse regression is first used to study the Gaussian graphical model,  and has been  extended to hypothesis testing problem in high dimensional linear regression \citep{liu2014hypothesis}. The procedure uses information from the precision matrix so that the correlations between test statistics become quantifiable. We combine this inverse regression procedure  with the estimation methods in \cite{lin2015regularization} to propose a test statistic with desired properties. In addition, in high dimensional setting, the sparsity assumption on the true regression coefficient results in a small number of alternatives, which leads to conservative false discovery rate (FDR) control. A less conservative approach  is to control the number of falsely discovered variables (FDV) \citep{liu2014hypothesis}. The proposed test statistic for single regression coefficient in IV models is shown to be asymptotically  normal and the proposed multiple testing procedure is shown  to control  the FDR or FDV. 

The remainder of the paper is organized as follows. Section
\ref{sec: method} presents  the high-dimensional IV model, the test statistics for single hypothesis and a  multiple testing procedure with the control of FDR or FDV.  Section \ref{sec: theory} provides the theoretical results of the single coefficient test statistic and the multiple testing procedure. Simulation results are presented  in Section \ref{sec:simulation}. An analysis of the yeast data set using proposed methods is given  in Section \ref{sec: application}. Discussion and suggestions for future work are provided in Section \ref{sec: discussion}. Proofs of the theorems are included as online Supplemental Materials.

\section{IV Models and Proposed Methodology} \label{sec: method}
The notations used in the paper are first given here. For any set $S$, $|S|$ denotes its cardinality. For a vector $x$, $\supp (x)$ is its support, $\|x\|_{p}$ is the standard $\ell_{p}$-norm and $\|x\|_0$ is defined as $|\supp (x)|$. For any matrix $A = (a_{ij})$, $i \in I, j \in J$ and subset $S \subset I, R \subset J$, $A_{S,R}$ denotes the submatrix $\{(a_{ij}): i \in S, j\in R \}$ and  $A_{-S,-R}$ denotes the submatrix $\{(a_{ij}): i \notin S, j\notin R \}$. For a matrix $A$,  $A_{\cdot, j}$  represents the $j$-th column of this matrix. For a sequence of random variables $x_n$ and a random variable $x$, $x_n \rightsquigarrow x$ implies  $x_n$ converges weakly to $x$ as $n \rightarrow \infty$.  Finally,   $a \wedge b$  represents the minimum value between $a$ and $b$, and  $a \lesssim b$ if there exists some constant $C$ such that $ a \leq Cb$ and $a \lesssim_p b$ if the inequality $ a \leq Cb$ holds with probability  going to 1.   

\subsection{Sparse Instrumental Variable Model}
Denote $Y \in \R^n$ as the $n$-dimension phenotype  vector, $\bX \in \R^{n \times p}$ as the gene expression matrix of $p$ genes and $\bZ \in \R^{n \times q}$ as the matrix of $q$ possible instrumental variables such as the genotypes of $q$ genetic variants.  \citet{lin2015regularization} considered the following high dimensional  IV regression model:
\begin{align}
Y & = \bX \bbeta_0 + \bm{\eta}, \label{eq:stage 2}\\
\bX & = \bZ \bGamma_0 + \bE, \label{eq:stage 1}
\end{align}
where $\bbeta_0 \in \R^{p}$ is the vector of regression coefficients that reflects the association between phenotype $Y$ and gene expression $\bX$, while $\bGamma_0$ reveals the relationships between the gene expressions  $\bX$ and the genetic variants $\bZ$. Without lose of generality, we assume $\bZ$ is centered and standardized. The error terms $\bmeta = (\eta_1, \eta_2, \ldots, \eta_n)^\top$ and $\bE = ( \bm{\varepsilon_1}, \ldots,\bm{\varepsilon_n})^\top$ are $n$-dimensional vector and $n$ by $p$ matrix, respectively. The joint distribution of $\left( \bm{\varepsilon_i}^\top, \eta_i \right)$ is a multivariate normal distribution with mean $0$, covariance matrix $\bSigma_{e}$ and is independent with $\bZ$. To emphasis the correlation between $Y$ and $\bX$, we assume that the correlation between $\bm{\varepsilon_i}$ and $\eta_i$ is not zero. In this paper we are interested in the high-dimensional setting where the dimension of the covariates $p$ and the dimension of potential instrumental variables  $q$ can  both be larger than $n$.  

As suggested by \citet{lin2015regularization}, estimation of $\bbeta_0$ in sparse setting can be performed by a two-stage penalized least squares method. To be specific, we first estimate the coefficients matrix $\bGamma_0$ in \eqref{eq:stage 1} column by column as the following:
\begin{align} \label{eq: gamma hat}
\wbGamma_{\cdot,j} = \argmin_{\bgamma \in \R^{q}}\left( \frac{1}{2n} \|\bX_{\cdot,j} - \bZ \bgamma \|_{2}^{2} + \lambda_{2j}\|\bgamma\|_1\right),  \quad j=1,2,\ldots,p,
\end{align}
where $ \lambda_{2j}$ is a tuning parameter. 
After obtaining an estimate of $\bGamma_0$, we plug in the predicted value of $\bX$, which is $\wbX = \bZ \widehat{\bGamma}$, to the second stage model \eqref{eq:stage 2} and obtain an estimator of $\bbeta_0$:
\begin{align} \label{eq: beta hat}
\widehat{\bbeta}= \argmin_{\bbeta \in \R^{p}}\left( \frac{1}{2n} \|Y - \wbX \bbeta \|_{2}^{2} + \lambda_{1}\|\bbeta\|_1\right),
\end{align}
where $\lambda_1$ is a tuning parameter. 

The focus of this paper is to  develop statistical test  of $\H_0:  \beta_{0i}=0$ for a given $i$  and to develop a procedure for the multiple hypothesis testing problem:
\begin{align*}
\H_{0i}: \ \beta_{0i} = 0 \quad \textrm{vs.} \quad \H_{1i}: \ \beta_{0i} \neq 0, \quad i=1,2,\ldots,p,
\end{align*}
with a correct control of FDR or FDV.  

\subsection{Hypothesis Testing for a Single Hypothesis Using Inverse Regression}
Denote $\bD = \bZ \bGamma_0$, from models \eqref{eq:stage 2} and \eqref{eq:stage 1}, 
\begin{align}
Y & =\mu +  \bD \bbeta_0 + \bm{\xi}, \label{eq: reduced form}
\end{align}
where $\bm{\xi} = \bmeta + \bE \bbeta_0$. When $\bZ$ consists of all the valid instruments, $\bD$ and  $\bm{\xi}$ are independent by the causal assumptions for a valid instrument and  \eqref{eq: reduced form} can be treated as a standard linear regression. 
Using the idea of inverse regression \citep{liu2014hypothesis,liu2013gaussian},  for each $i=1,2, \ldots, p$,  $\bD_i$ is regressed  on $(Y, \bD_{\cdot, -i})$ as:
\begin{align} 
\bD_{\cdot, i} = a_i + \left(Y, \bD_{\cdot, -i} \right)\btheta_i + \bzeta_i,   \label{eq: inverse regression}
\end{align}
where  $\bzeta_i$ satisfies $\E \bzeta_i = 0$ and is uncorrelated with $(Y, \bD_{\cdot, -i})$. Based on the properties of multivariate normal distribution \citep{anderson2003introduction}, the regression coefficient $\btheta_i$ is related to  the target parameter $\bbeta_0$ by the following equality:
\begin{align} \label{eq: theta in inverse regression} 
\btheta_i = -\sigma^2_{\zeta_i} \left( -\frac{\beta_{0i}}{\sigma^2_{\xi}},\frac{\beta_{0i}\bbeta_{-0i}^{\top}}{\sigma^2_{\xi}} + \bOmega_{-i,i}^{\bD} \right),
\end{align}
where $\sigma^2_{\zeta_i}$ and $\sigma^2_{\xi}$ denote the variance of $\zeta_i$ and $\xi$, respectively, and $\bOmega^{\bD} = \bSigma^{-1}_{\bD}$ is the precision matrix for $\bD$. Since $\cov(\bD, \xi) =0$, we have  $\sigma^2_{\zeta_i}\beta_{0i} = \sigma^2_{\xi}\btheta_{i1} = \btheta_{i1}\cov (\xi, y) = -  \cov(\xi,\zeta_i)$, therefore, the null hypothesis $\H_{0i}: \beta_{0i}=0$ is  equivalent to 
\begin{align*}
\H_{0i}: \ \cov(\xi,\zeta_i) = 0 \quad \textrm{vs.} \quad \H_{1i}: \ \cov(\xi,\zeta_i) \neq 0, \quad i=1,2,\ldots,p. 
\end{align*}

Since the data observed are $\{y_k, \bX_k, \bZ_k, k=1,2,\cdots n\}$,  the vector $\bD_{ i}$ in \eqref{eq: inverse regression} is not  observed for any $i=1,2,\ldots,p$.  One can   estimate $\btheta_i$ via regularization by replacing $\bD$ with its estimated value $\wbD = \wbX = \bZ \widehat{\bGamma}$, 
\begin{align}  \label{eq: theta hat}
\wbtheta_i = \argmin_{\btheta} \left\{\frac{1}{2n} \|\wbD_{\cdot, i} - \left(Y, \wbD_{\cdot, -i} \right)\btheta_i  \|_{2}^{2} + \mu_i\|\btheta\|_1 \right\}, \quad i=1,2,\ldots,p,
\end{align}
where $\mu_i$ is a tuning parameter. 

The sample correlation between $\xi$ and $\zeta_i$ is then used to construct the test statistic for $\H_{0i}$ \citep{liu2013gaussian}. Using the estimates $\wbbeta$, $\wbD$ and $\wbtheta_i$, the estimated residuals are 
\begin{align*}
\widehat{\xi}_k  &= y_k - \overline{Y} - \left( \wbD_k - \overline{\wbD} \right)^\top \wbbeta,   \\
\widehat{\zeta}_{k,i} &= \wbD_{k,i} - \overline{\wbD_i} - \left(y_k - \overline{Y}, \left( \wbD_{k, -i}- \overline{\wbD}_{-i}\right)^\top\right) \wbtheta_i,
\end{align*}
for $k=1,2,\ldots,n$ and $i=1,2,\ldots,p$, where
\begin{align*}
\overline{Y}  =  \dfrac{1}{n} \sum_{k=1}^{n} y_k, \  \overline{\wbD}  = \dfrac{1}{n} \sum_{k=1}^{n} \wbD_k , \ \overline{\wbD_i}  =  \dfrac{1}{n} \sum_{k=1}^{n} \wbD_{k,i},
\  \overline{\wbD}_{-i}  = \dfrac{1}{n} \sum_{k=1}^{n} \wbD_{k, -i}.
\end{align*}
Using the bias correction formula in \citet{liu2013gaussian}, for each $i$, define the test statistic  as  
\begin{align*}
T_i = \sqrt{n}\left(\frac{1}{n} \sum_{k=1}^{n} \widehat{\xi}_k \widehat{\zeta}_{k,i} + \frac{1}{n} \sum_{k=1}^{n} \widehat{\xi}_k^2 \wbtheta_{1,i} + \frac{1}{n} \sum_{k=1}^{n}  \widehat{\zeta}_{k,i}^2 \wbbeta_i \right) \bigg / \widehat{\sigma}_{\xi}\widehat{\sigma}_{\zeta_i},
\end{align*}
where
\begin{align*}
\widehat{\sigma}^2_{\xi} = \frac{1}{n}\sum_{k=1}^{n} \widehat{\xi}_k^2, \quad \widehat{\sigma}^2_{\zeta_i} = \frac{1}{n}\sum_{k=1}^{n} \widehat{\zeta}_{k,i}^2.
\end{align*}
The bias correction formula adds two extra terms to the original sample correlation in order to eliminate the higher order bias  resulting from  the bias of the Lasso-type estimator. Using the transformation theorem in \citet{anderson2003introduction}, the final test statistic for testing $\H_{0i}: \ \cov(\xi,\zeta_i) = 0$ is  defined as 
\begin{align*}
\widehat{T}_i = \frac{T_i}{1 - \frac{T_i^2}{n}\bm{1}\left(\frac{T_i^2}{n}<1\right)}, 
\end{align*} 
which has an asymptotic $N(0,1)$ distribution under the null (see Theorem 1).

\subsection{Rejection Regions for Multiple Testing Procedure with FDR and FDV control}
After obtaining the test statistic $\widehat{T}_i $ for $\H_{0i}$, we determine the rejection region for simultaneous tests of  $\widehat{T}_i $ for $\H_{0i}$ for $i=1,\cdots, p$.  Recall that the definitions of FDR and FDV are:
\begin{align*}
FDR  = \E \left\{   \dfrac{ \sum_{i \in \H_0}\bm{1} \left( |\widehat{T}_i | \geq t\right) }{\sum_{i=1}^{p}\bm{1}\left( |\widehat{T}_i | \geq t\right)\lor 1}\right\},  FDV  = \E \left\{  \sum_{i \in \H_0}\bm{1} \left( |\widehat{T}_i | \geq t\right) \right\}.
\end{align*}
Suppose the rejection region for each $\H_{0i}$ is \{$|\widehat{T}_i | \geq t $\},  by the definition of false discovery proportion and false discovery rate, an ideal choice of $t$ that  controls the FDR below a certain level $\alpha$ is
\begin{align*}
t_0 = \inf \left\{0 \leq t \leq \sqrt{2 \log p }: \dfrac{\sum_{i \in \H_0} \bm{1}\left( |\widehat{T}_i | \geq t\right)}{\sum_{i=1}^{p}\bm{1}\left( |\widehat{T}_i | \geq t\right)\lor 1 } \leq \alpha \right\}.
\end{align*}

In practice the quantity $\sum_{i \in \H_0} \bm{1}\left( |\widehat{T}_i | \geq t\right)$ can be estimated by $2p\left(1 - \Phi(t)\right)$, where $\Phi(t)$ is the cumulative distribution function of the standard normal distribution. Based on this approximation, the quantity $t_0$ in the multiple testing procedure can be estimated by
\begin{align} \label{eq: t_0}
\widehat{t}_0 = \inf \left\{0 \leq t \leq \sqrt{2 \log p }: \dfrac{2p\left(1 - \Phi(t)\right)}{\sum_{i=1}^{p}\bm{1}\left( |\widehat{T}_i | \geq t\right)\lor 1 } \leq \alpha \right\}.
\end{align} 
We reject the hypothesis $\H_{0i}$ if $|\widehat{T}_i| \geq \widehat{t}_0$ for $i=1,2, \ldots,p$. 

Similarly, to control the FDV, the rejection region $|\widehat{T}_i| \geq \widehat{t}_0$ is given by
\begin{align} \label{eq: t_0 FDV}
\widehat{t}_0 = G^{-1}\left(\dfrac{k}{p}\right),
\end{align} 
where $G(t)= 2(1 - \Phi(t))$.

\subsection{Implementation}
The construction of the test statistics involves a set of convex optimizations and selection of the tuning parameters in order to solve the Lasso regressions \eqref{eq: gamma hat}, \eqref{eq: beta hat} and \eqref{eq: theta hat}. The optimizations can be  efficiently implemented using the  coordinate descent (CD) algorithm \citep{friedman2010regularization,lin2015regularization}.  The CD algorithm is a well-known and widely used convex optimization algorithm for penalized regressions so we omit the details here. 

For tuning parameter selection, we have separate strategies for the two groups of tuning parameters $\lambda$ and $\mu$. For the optimization problems \eqref{eq: gamma hat} and \eqref{eq: beta hat}, the tuning parameters $\lambda_{1}$ and $\lambda_{2j}$, $j=1,2,\ldots,p$ can be chosen by a $K$-fold cross-validation (CV) for $K=5$ or $10$, where $\lambda^{\text{opt}}_{1}$ and $\lambda^{\text{opt}}_{2j}$, $j=1,2,\ldots,p$ are determined by minimizing the CV errors of the corresponding optimization problem. When both $p$ and $q$ are very large, performing CV can  be time-consuming. So in our simulations and real data applications, we applied an alternative method for selecting these two groups of tuning parameters that relies on scaled Lasso \citep{sun2012scaled},  which is computationally more efficient.

Selection of the tuning parameters for the inverse regression \eqref{eq: theta hat} is done by  a data-driven procedure as  suggested by \citet{liu2013gaussian} and \citet{liu2014hypothesis}. To be specific, let $\delta_j = j$ for $j=1,2,\ldots, 100$ and $\mu_{j} = 0.02\delta_j \sqrt{\widehat{\bSigma}^{D}_{i,i} \log p / n}$, where $\widehat{\bSigma}^{D}$ is the sample covariance matrix of $\wbD$. The choice of the $\delta$ is determined by:
\begin{align*}
\hat{\delta} = \argmin_{\delta}\sum_{k=30}^{90}\left\{\dfrac{\sum_{i=1}^{p}\bm{1}\left( |\widehat{T}_i| \geq \Phi^{-1}\left(1 - k/200\right)\right)}{kp/100}-1\right\}^2.
\end{align*}
The tuning parameter $\mu_i$ in \eqref{eq: theta hat} is chosen as $\hat{\mu}_i = 0.02\hat{\delta} \sqrt{\widehat{\bSigma}^{D}_{i,i} \log p / n}$.

\section{Theoretical Results} \label{sec: theory}
We provide in this section some theoretical results  of the proposed methods. We first restate  the estimation error bounds of $\bGamma_0$ and $\bbeta_0$  in models \eqref{eq:stage 2} and \eqref{eq:stage 1} derived in  \citet{lin2015regularization}, which are needed in  constructing the test statistics. 
Before stating the results, we first introduce some assumptions. 
For any matrix $\bX$, we say it satisfies the restricted eigenvalue (RE) condition if its restricted eigenvalue is strictly bounded away from $0$. That is, for some $1\leq s \leq p$, the following condition holds:
\begin{align*}
\kappa(s,\bX)  \triangleq \min_{ \substack{J \subseteq \{ 1,\ldots,p\}\\|J| \leq s}} \min_ {\substack{\bdelta \neq 0 \\ \|\bdelta_{J^c}\|_1 \leq  3\|\bdelta_{J}\|_1}} \dfrac{\|\bX \bdelta\|_2}{\sqrt{n}\|\bdelta_{J}\|_2} >0.
\end{align*} 
Denote $s_1 = \|\bbeta_0\|_0$, $s_2 = \max_{j} \|\bGamma_{\cdot,j}\|_0$, $r = \max_{j} \|\btheta_j\|_0$ and $\kappa$ is the restricted eigenvalue defined above. The following assumptions are needed:
\begin{enumerate}
	\item [(A1)] The instrumental variable matrix $\bZ$ and matrix $\bD = \bZ \bGamma_0$ satisfies the restricted eigenvalue condition with some constants $\kappa(s_2, \bZ) , \kappa(s_1, \bD)>0$, respectively. 
	\item [(A2)]  There exists a positive constant $C$ such that $\max \{ \|\bbeta_0\|_1, \|\bGamma_0\|_1,  \{\|\btheta_i\|_1\}_{i=1,\ldots,p}\}\leq C$.
	\item [(A3)] There exists a positive constant $C$ such that $\max_{1 \leq j \leq p} \left( \bSigma_{j,j}^{e} \right) \leq C^2$. 
	\item [(B1)] In the inverse regression model \eqref{eq: inverse regression}, denote $\bM_i = \left(Y, \bD_{\cdot, -i} \right)$, for $i=1,\ldots,p$, then $\bM_i$ satisfies the restricted eigenvalue condition with some constant $\kappa(r, \bM_i)$. In addition, assume that there exists a positive constant $\kappa(Y, \bD)$ such that $\min_{i} \kappa(r, \bM_i) \geq \kappa(Y, \bD)$.
	\item [(C1)] The precision matrix $\bOmega^{\bD}$ and covariance matrix $\Sigma_{\bD}$ satisfies $\max_{1 \leq j \leq p} \left( \bOmega_{j,j}^{\bD}, \Sigma_{j,j}^{\bD}\right) \leq C$ for some constant $C$ and $\var (Y_i) \leq C$.
	\item [(C2)] The dimensional parameters  $n,p,q,s_1,s_2,r$ satisfy the following asymptotic scaling condition as $n\rightarrow \infty$:
	$$
	\max\{r \sqrt{s_2},s_1,s_2 \} \sqrt{\dfrac{\log p \left(\log p + \log q\right)}{n}} = o(1).
	$$
	\item [(C3)] The precision matrix $\bOmega^{\bD}$ satisfies the following condition: for some $\varepsilon >0$ and $\delta>0$,
	$$\sum_{(i,j) \in \mathcal{A}(\varepsilon)}p^{\frac{2|\rho_{ij,\omega_{\bD}}|}{1 + |\rho_{ij,\omega_{\bD}}|}+\delta} = \mathcal{O}(p^2/(\log p)^2),$$ where $\rho_{ij,\omega_{\bD}} = \Omega_{ij}^{\bD}/(\Omega_{ii}^{\bD}\Omega_{jj}^{\bD})^{1/2}$ and $\mathcal{A}(\varepsilon) = \mathcal{B}((\log p)^{-2-\varepsilon})$ with $\mathcal{B}(\delta) = \{(i,j): |\rho_{ij,\omega_{\bD}}|\geq \delta, i\neq j \}$.
\end{enumerate}

These assumptions play different roles in establishing the asymptotic results. To be specific, assumptions (A1) to (A3) are required to obtain the estimation error bounds for $\wbbeta$ and $\wbGamma_{\cdot,j}$. These assumptions are similar to those in \citet{bickel2009simultaneous} and are used in \citet{lin2015regularization}.  They require that matrix $\bZ$ and $\bD$ are well-behaved and $\ell_1$ norms of the true parameters $\bbeta_0$, $\bGamma_0$ are bounded away from infinity. Assumption (B1) guarantees that $\btheta_i$ can be well estimated. This assumption is implicitly assumed, though not stated, in \citet{liu2014hypothesis}. Assumptions (C1) and (C2) are needed to obtain the asymptotic distribution of $\widehat{T}_i$. Particularly, assumption (C1) bounds the entries of the covariance matrix $\Sigma_{\bD}$ and precision matrix $\Omega^{\bD}$ and assumption (C2) provides the relation among the dimension and sparsity  parameters  $n,p,q,s_1,s_2$ and $r$, where  $s_1$, $s_2$ and $r$ control the sparsity of $\bbeta_0$, $\bGamma_0$ and $\btheta_{i}$ respectively.  Assumption (C3) is used for controlling the FDR, which imposes  some conditions on the precision matrix  \citep{liu2014hypothesis}. In addition, if we fix $q$, which is the number of instruments, then assumption (C2) is equivalent to $\log p = o(\sqrt{n})$. This assumption is often made in the inference results related with Lasso and other high dimensional models \citep{gold2017inference,javanmard2014confidence,ning2017}.

\subsection{Asymptotic distribution of test statistic for single null hypothesis}
Since our test statistics rely on the estimation of the parameters in  models \eqref{eq:stage 2} and \eqref{eq:stage 1}, we first provide a lemma on the estimation errors of $\bGamma_{\cdot, j}$ and $\bbeta$.
\begin{lemma}[Estimation error bounds of $\bGamma_{\cdot, j}$ and $\bbeta_0$ \citep{lin2015regularization}] \label{lemma: gamma and beta}
	Under assumptions (A1)-(A3), for each $j=1,2, \ldots, p$, if the tuning parameter $\lambda_{2j}$ is chosen as
	\begin{align*}
	\lambda_{2j}  = \widetilde{C} \sqrt{\dfrac{\bSigma^{e}_{j,j} \left(\log p + \log q\right)}{n}},
	\end{align*}
	for some $\widetilde{C} \geq 2\sqrt{2}$, then with probability at least $1 - \left(pq\right)^{1 - \widetilde{C}^2/8}$, $\widehat{\bGamma}$ defined in \eqref{eq: gamma hat} satisfies
	\begin{align*}
	\|\widehat{\bGamma} - \bGamma_0 \|_1 \leq \dfrac{16\widetilde{C}C}{\kappa^2(s_2,\bZ)}s_2\sqrt{\dfrac{\log p + \log q}{n}},
	\end{align*}
	and 
	\begin{align*}
	\|\bZ \left( \wbGamma - \bGamma_0\right)\|_{F}^2 \leq \dfrac{16\widetilde{C}^2C^2}{\kappa^2(s_2,\bZ)}s_2p \left( \log p + \log q\right). 
	\end{align*}
	Furthermore, if the set of tuning parameters $\{\lambda_{2j}: j=1,\ldots,p\}$ satisfy
	\begin{align*}
	\lambda_{\max} (2C + \lambda_{\max}) \leq \dfrac{\kappa^2(s_2,\bZ)\kappa^2(s_1, \bD)}{1024s_1s_2},
	\end{align*}
	where $\lambda_{\max}  = \max_{1 \leq j \leq p} \lambda_{2j}$, if $\lambda_{1}$ is chosen as:
	\begin{align*}
	\lambda_{1}  = C_0 \sqrt{\dfrac{s_2\left(\log p + \log q\right)}{n}},
	\end{align*}
	then with probability at least $1 - C_1(pq)^{-C_2}$, $\widehat{\bbeta}$ defined in \eqref{eq: beta hat} satisfies
	\begin{align*}
	\|\widehat{\bbeta}  - \bbeta_0 \|_1 \leq C_3s_1\sqrt{\dfrac{s_2\left(\log p + \log q\right)}{n}},
	\end{align*}
	for some positive constants $C_0-C_3$. 
\end{lemma}
In addition, we have the following lemma on the estimation error bound of $\btheta_i$. 
\begin{lemma}[Estimation error bounds of $\btheta_i$] \label{lemma: theta}
	Under assumptions (A1)-(A3) and (B1), for each $i=1,2,\ldots,p$, there exists some positive constants $C_4, C_5,C_5^{*}$, if the tuning parameter $\mu_i$ is chosen as
	\begin{align*}
	\mu_i =  \dfrac{C_4^{*}}{\kappa(s_2,\bZ)}\sqrt{\dfrac{s_2(\log p + \log q)}{n}},
	\end{align*}
	with $C_4^{*} = C_5^{*}\max(C, \sigma_{\zeta_i})$,
	then with probability at least $1 - C_4\left(pq\right)^{-C_5}$, $\wbtheta_i$ in \eqref{eq: theta hat} satisfies
	\begin{align*}
	\|\wbtheta_i  - \btheta_i \|_1 \leq \dfrac{64C_4^{*}}{\kappa^2(Y, \bD)\kappa(s_2,\bZ)}r\sqrt{\dfrac{s_2(\log p + \log q)}{n}}.
	\end{align*}
\end{lemma}

Based on Lemmas \ref{lemma: gamma and beta} and \ref{lemma: theta}, the following theorem provides the asymptotic distribution of the test statistic $\widehat{T}_{i}$ under  the null $\H_{0i}$. 
\begin{theorem}[Asymptotic distribution of $\widehat{T}_{i}$] \label{thm: T_i}
	Under assumptions (A1)-(A3), (B1) and (C1)-(C2), with the proper choices of the tuning parameters $\lambda_{1}$, $\lambda_{2j}$ and $\mu$ as stated in Lemma \ref{lemma: gamma and beta} and \ref{lemma: theta},  for  each $i=1,2.\ldots,p$, under the null $\H_{0i}: \beta_{0i} =0$, 
	\begin{align*}
	\widehat{T}_{i} \rightsquigarrow N(0,1).
	\end{align*}
\end{theorem}
This null distribution can be used to test the individual null hypothesis $\H_{0i}: \beta_{0i} =0$.

\subsection{Theoretical results on FDR and FDV}
The next theorem shows that the proposed multiple testing procedure controls the FDR.  
\begin{theorem}[Asymptotic result for multiple testing procedure]\label{thm: FDR}
	Denote FDR$ = \textrm{FDR}(\widehat{t}_0)$, assuming (A1)-(A3), (B1) and (C1), (C3) hold, $p \leq n^c$ for some $c>0$. We further assume a condition stronger than $C2$ such as  the quantities in the left of assumption $C2$ are of order $o((\log p)^{-\frac{1}{2}})$ instead of $o(1)$, and for some $\widetilde{c}>2$, 
	\begin{align} \label{condition:infinte true alternatives}
	\sum_{i \in \mathcal{H}_1} \bm{1}\left(\dfrac{\bbeta_i}{\sqrt{\sigma^2_{\xi}\Omega_{i,i}^{\bD}}} \geq \sqrt{\widetilde{c} \log p/n}\right) \rightarrow \infty,
	\end{align}
	as $(n,p) \rightarrow \infty$. Then with the proper choice of all tuning parameters and the threshold $\widehat{t}_0$, with a pre-specified level $\alpha$, we have
	\begin{align*}
	\lim_{n,p \rightarrow \infty} \dfrac{\textrm{FDR}}{\alpha p_0 / p} =1.
	\end{align*}
\end{theorem}
This theorem indicates that under proper conditions, the empirical FDR is  controlled under a pre-specified level. Notice that in addition to the assumptions previously mentioned, we require a stronger condition \eqref{condition:infinte true alternatives}. This condition indicates that the number of true alternatives needs to tend to infinity, which is also  required   in \citet{liu2014hypothesis}. 

Similar to the result of the FDR but with weaker assumptions, for the FDV control, we have the following result:
\begin{theorem}[Asymptotic results for multiple testing procedure]\label{thm: FDV}
	Assuming (A1)-(A3), (B1) and (C1) hold, $p \leq n^c$ for some $c>0$ and we further assume a condition stronger than $C2$ such as  the quantities in the left of assumption $C2$ are of order $o((\log p)^{-\frac{1}{2}})$ instead of $o(1)$. Then with the proper choice of all tuning parameters and the threshold $\widehat{t}_0$, with a pre-specified level $k$, we have:
	\begin{align}
	\lim_{n,p \rightarrow \infty} \dfrac{\textrm{FDV}}{k p_0 / p} =1.
	\end{align}
\end{theorem}
Here to control the FDV, we do not need assumption (C3) on the precision matrix and condition \eqref{condition:infinte true alternatives}.

\section{Simulations} \label{sec:simulation}
We evaluate the performance of the proposed  methods through a set of  simulations. Following models \eqref{eq:stage 2} and \eqref{eq:stage 1}, we first generate the instruments matrix $\bZ$ where $\bZ_i \sim N(0, \bSigma_z)$. The covariance matrix $\bSigma_z$ satisfies $\left( \bSigma_z\right)_{ij} = 0.5^{|i-j|}$. For each $\bGamma_{\cdot, j}$, we first randomly pick $s_2$ out of $q$ nonzero entries and then each entry is generated randomly from a uniform distribution $U([-b,-a] \cup [a,b])$ with $a=0.75,b=1$. Parameter  $\bbeta_0$ is generated similarly where we pick $s_1$ out of $p$ nonzero entries and each entry is generated randomly from $U([-0.3,0.1] \cup [0.1,0.3])$. As for the joint distribution of $\left( \bm{\varepsilon_i}^\top, \eta_i \right)$, its covariance matrix $\bSigma_{e}$ is generate by: $\left(\bSigma_{e}\right)_{ij} = 0.2^{|i-j|}$ for $1\leq i,j\leq p$, $\left(\bSigma_{e}\right)_{p+1,p+1}=1$ and among $\left(\bSigma_{e}\right)_{i,p+1}$ where $i=1,\ldots,p$, 10 entires are picked randomly and set to be 0.3. We impose this structure so that $\eta_i$ is correlated with $\bm{\varepsilon_i}$.  

Covariates $\bX$ and response $Y$ are generated based on our model. We consider different values of $(n,p,q)$ with $(n,p,q) = (200,100,100), (400,200,\allowbreak 200), (200,500,500)$ and $(s_1,s_2) = (10,10)$. We compare our methods with the test developed  in \citet{liu2014hypothesis} for high dimensional regression analysis linking $Y$ to $\bX$ ignoring the fact that $\bX$ and $\bmeta$ are correlated.  It should be noted that the independent error assumption is necessary for the method in \citet{liu2014hypothesis} to work.

\subsection{Test of  Single Hypothesis}
\begin{figure}
	\centering
	\begin{tabular}{ccc}
		\includegraphics[height=0.15\textheight,width=0.33\textwidth]{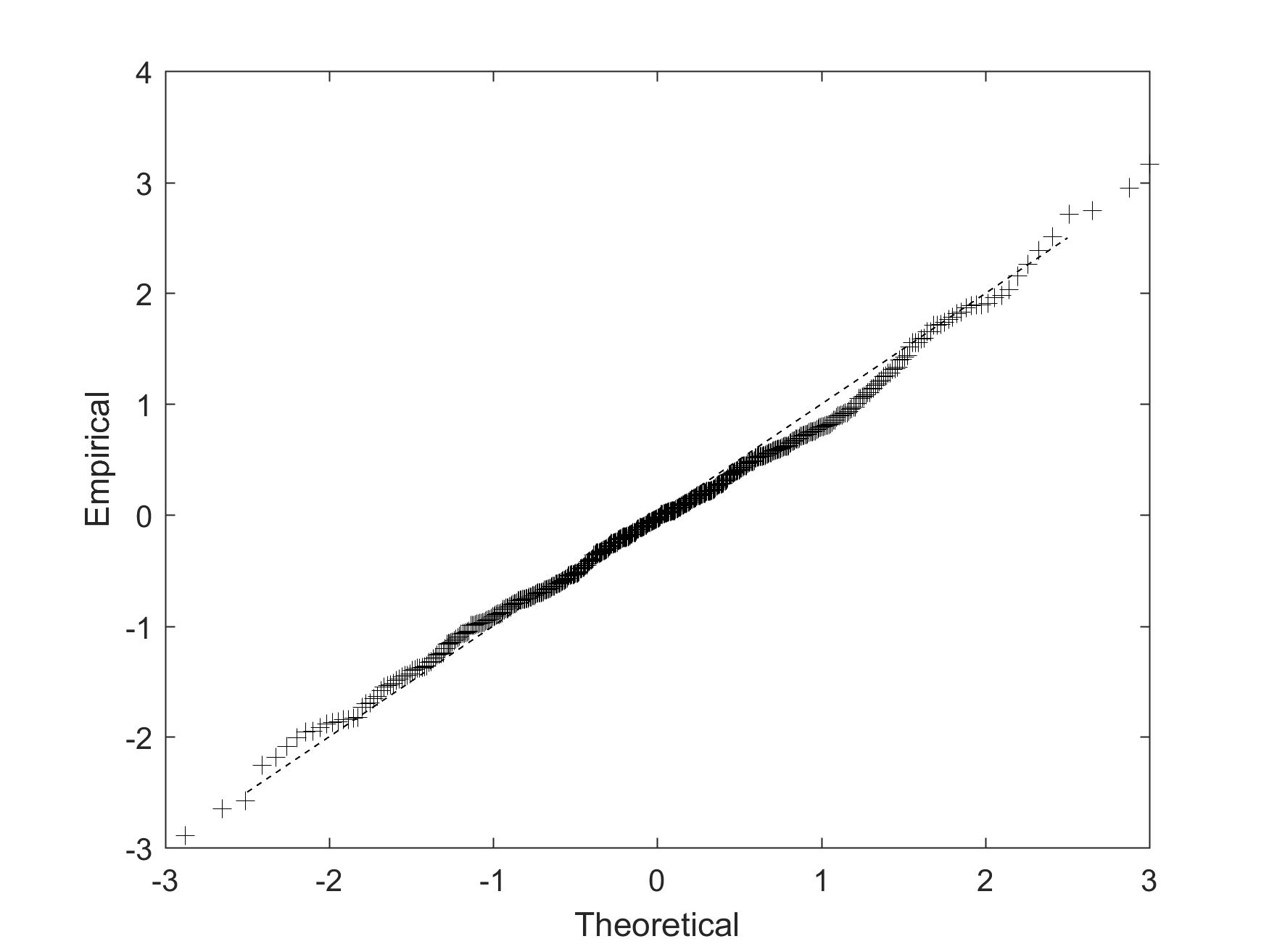} &	\includegraphics[height=0.15\textheight,width=0.33\textwidth]{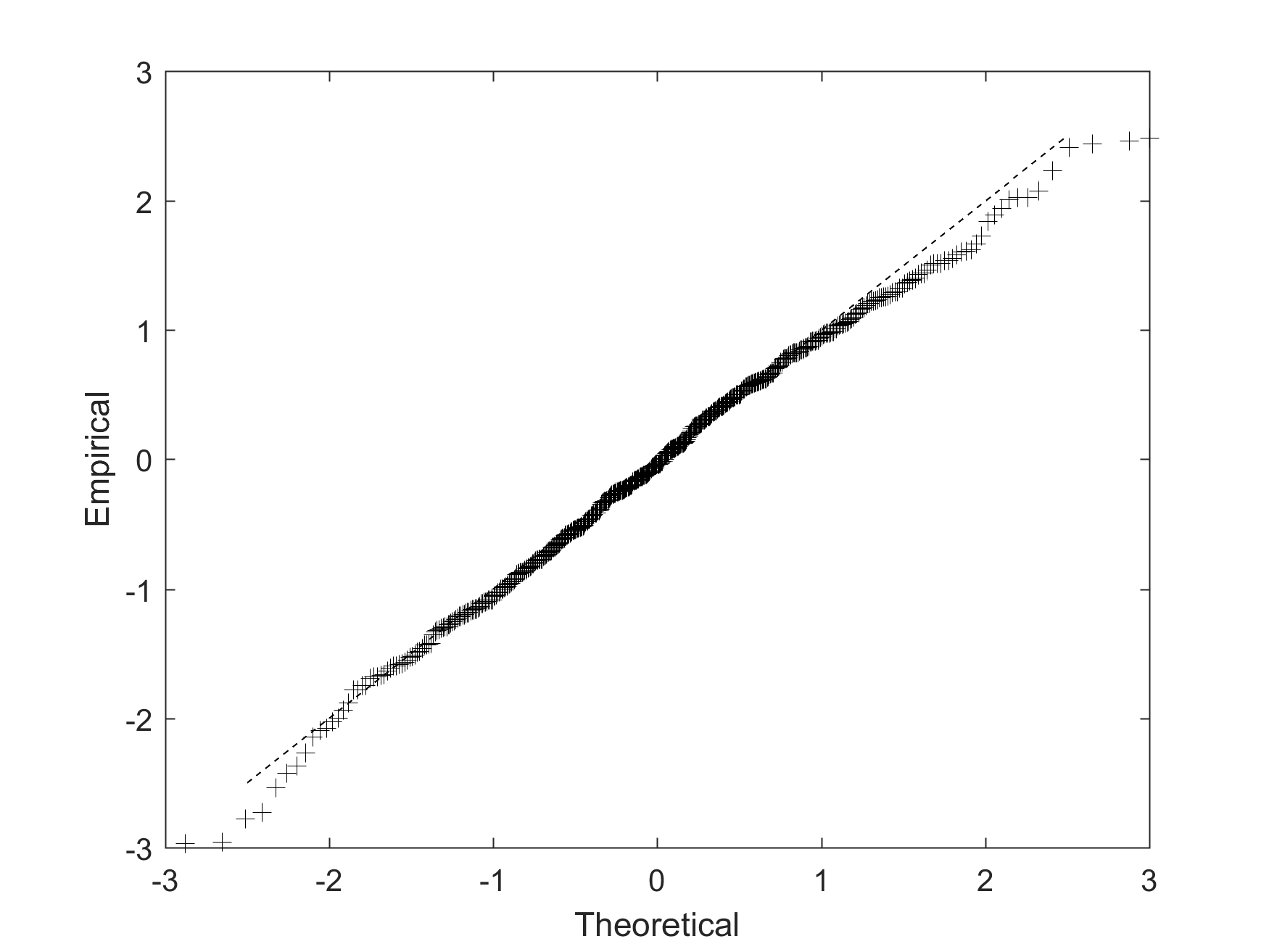} & \includegraphics[height=0.15\textheight,width=0.33\textwidth]{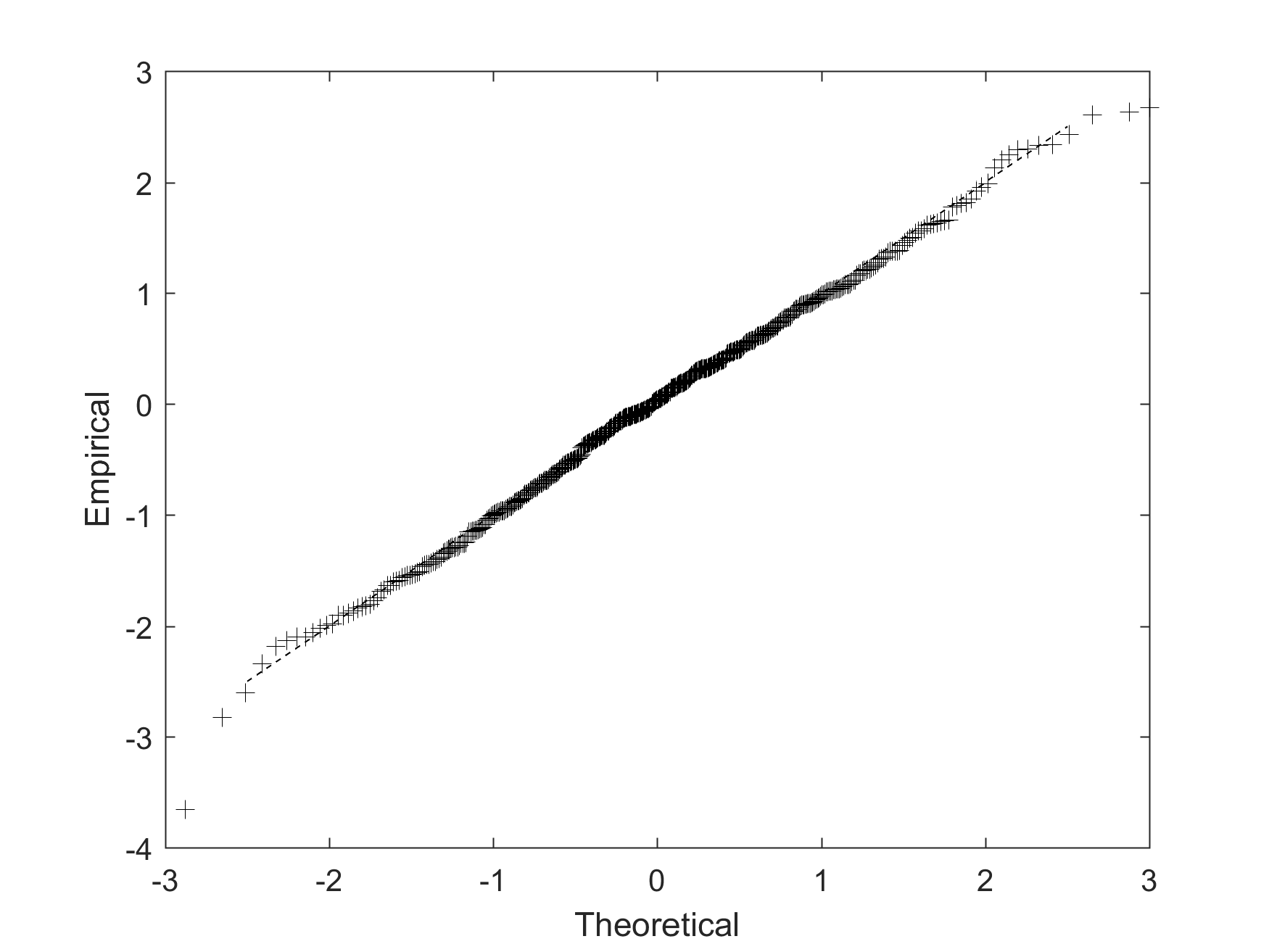} \\
		(a) & (b) & (c) \\
		\includegraphics[height=0.15\textheight,width=0.33\textwidth]{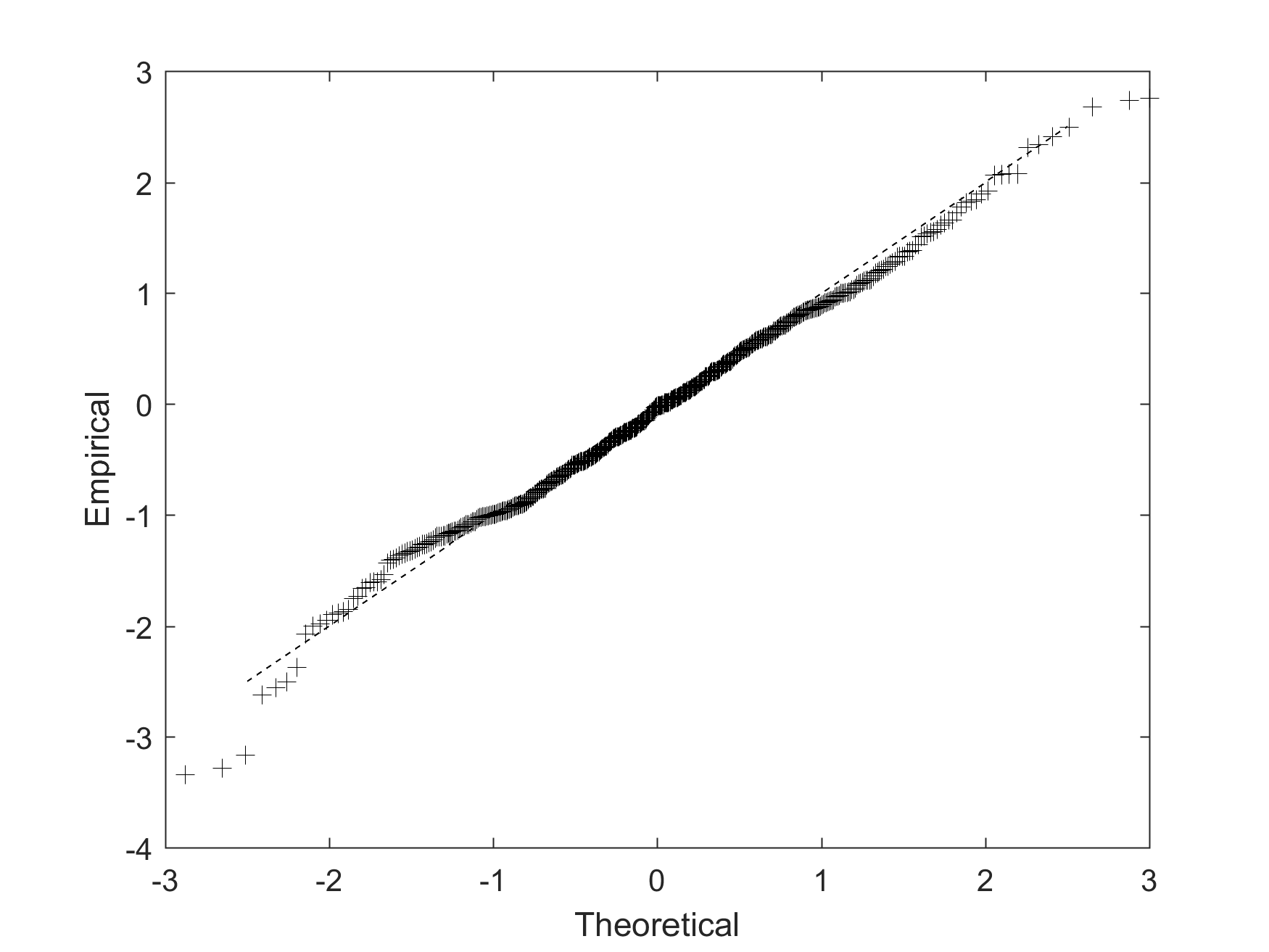} &	\includegraphics[height=0.15\textheight,width=0.33\textwidth]{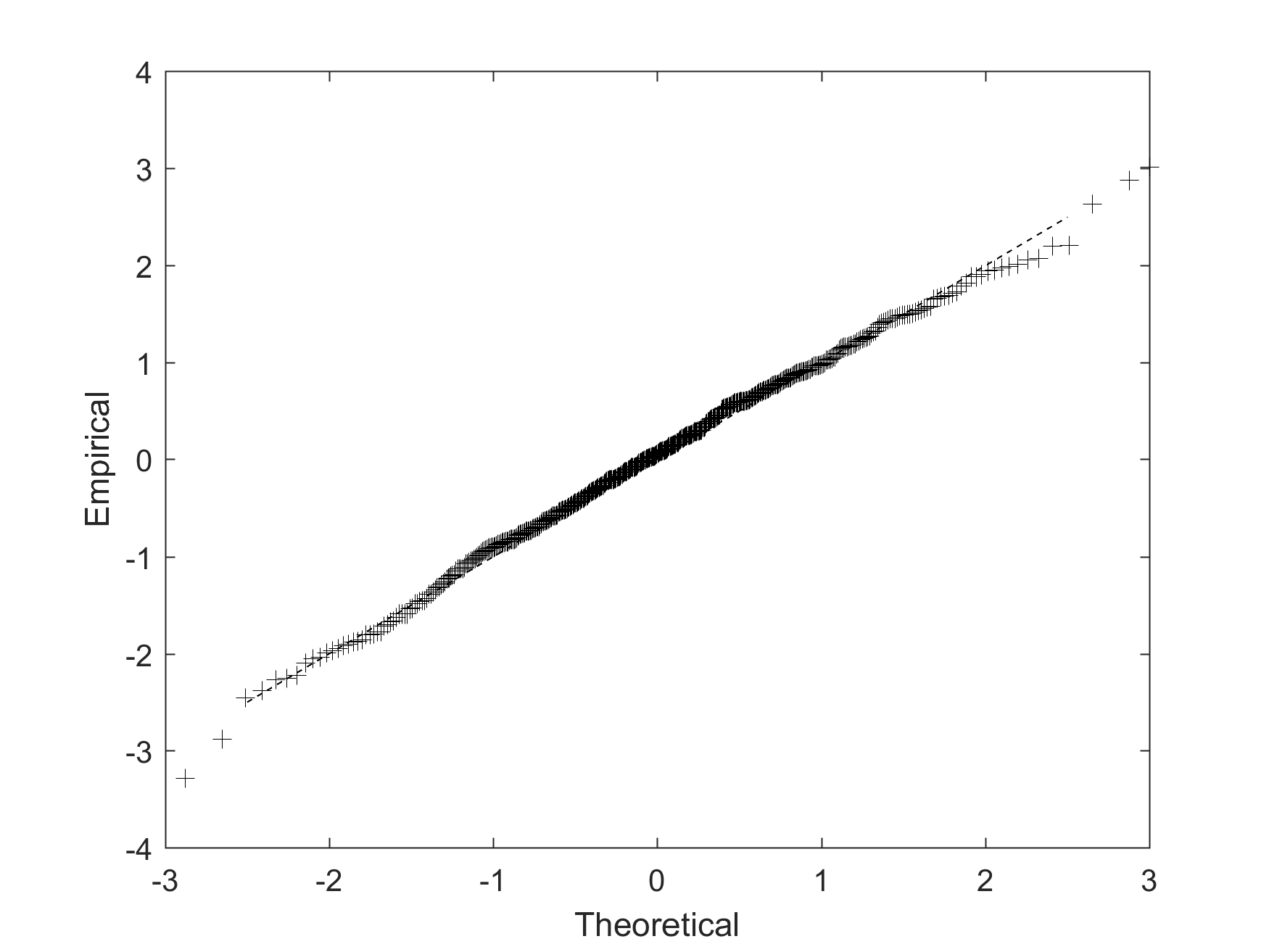} & \includegraphics[height=0.15\textheight,width=0.33\textwidth]{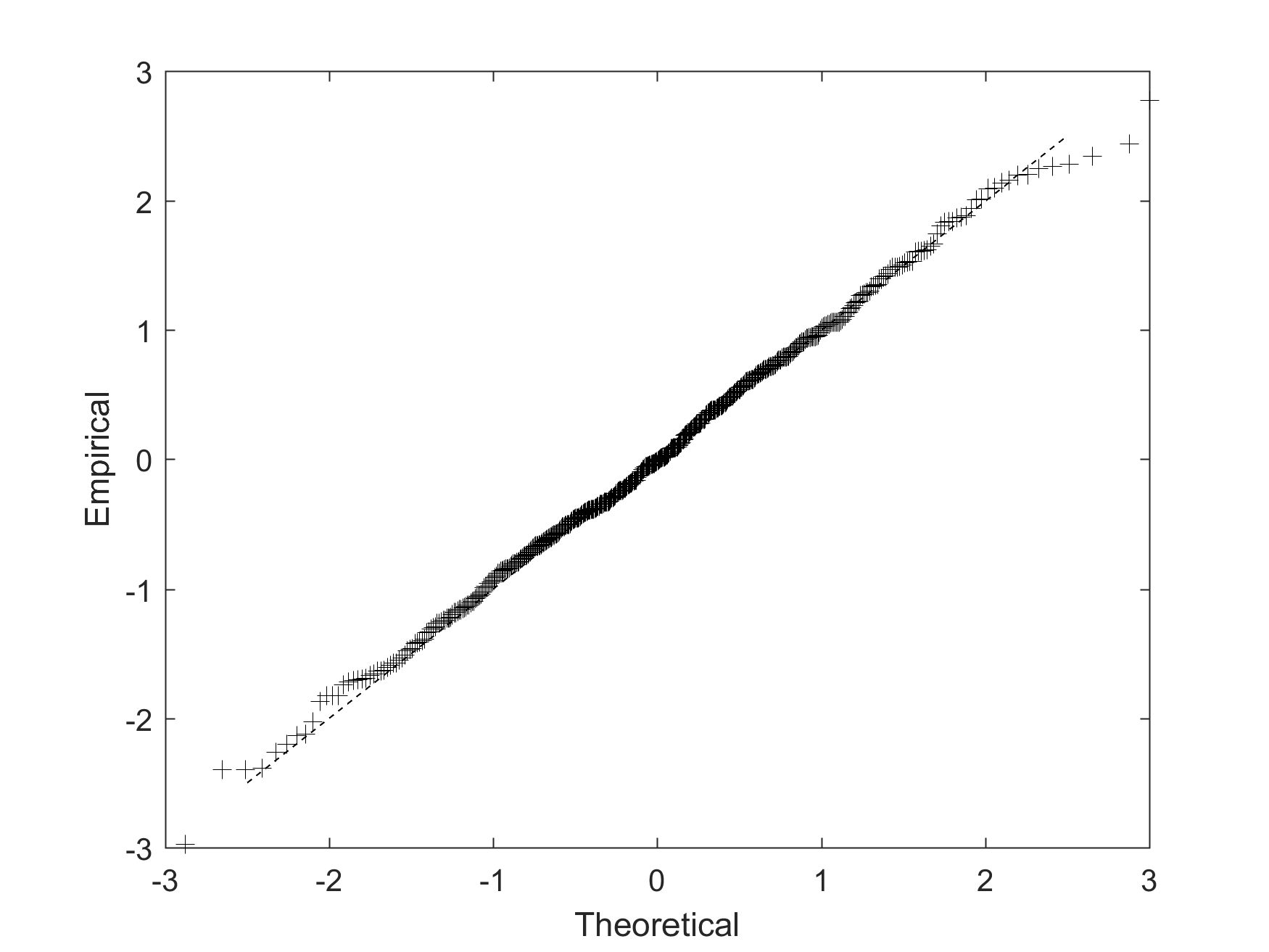} \\
		(d) & (e) & (f) \\
		\includegraphics[height=0.15\textheight,width=0.33\textwidth]{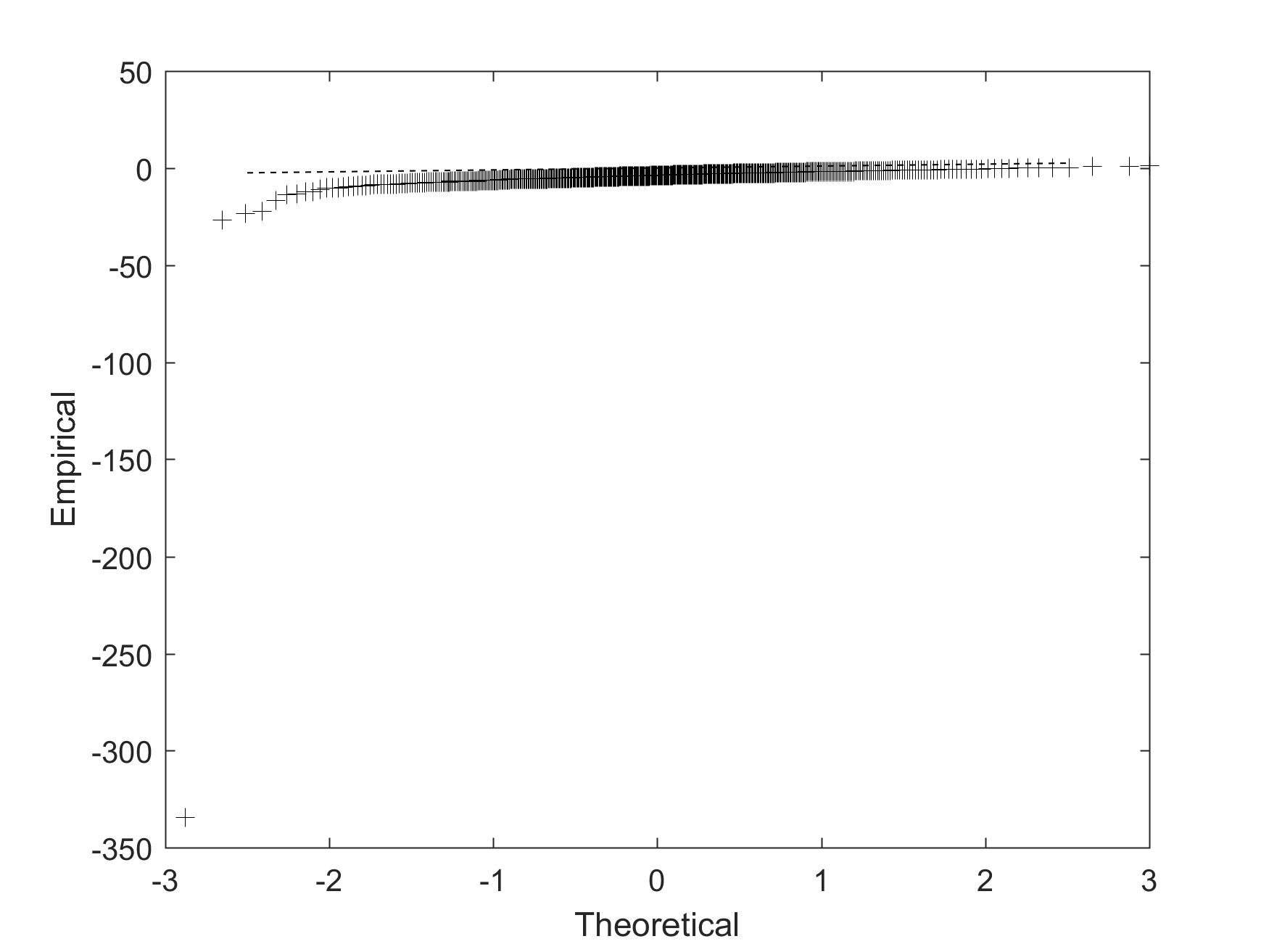} &	\includegraphics[height=0.15\textheight,width=0.33\textwidth]{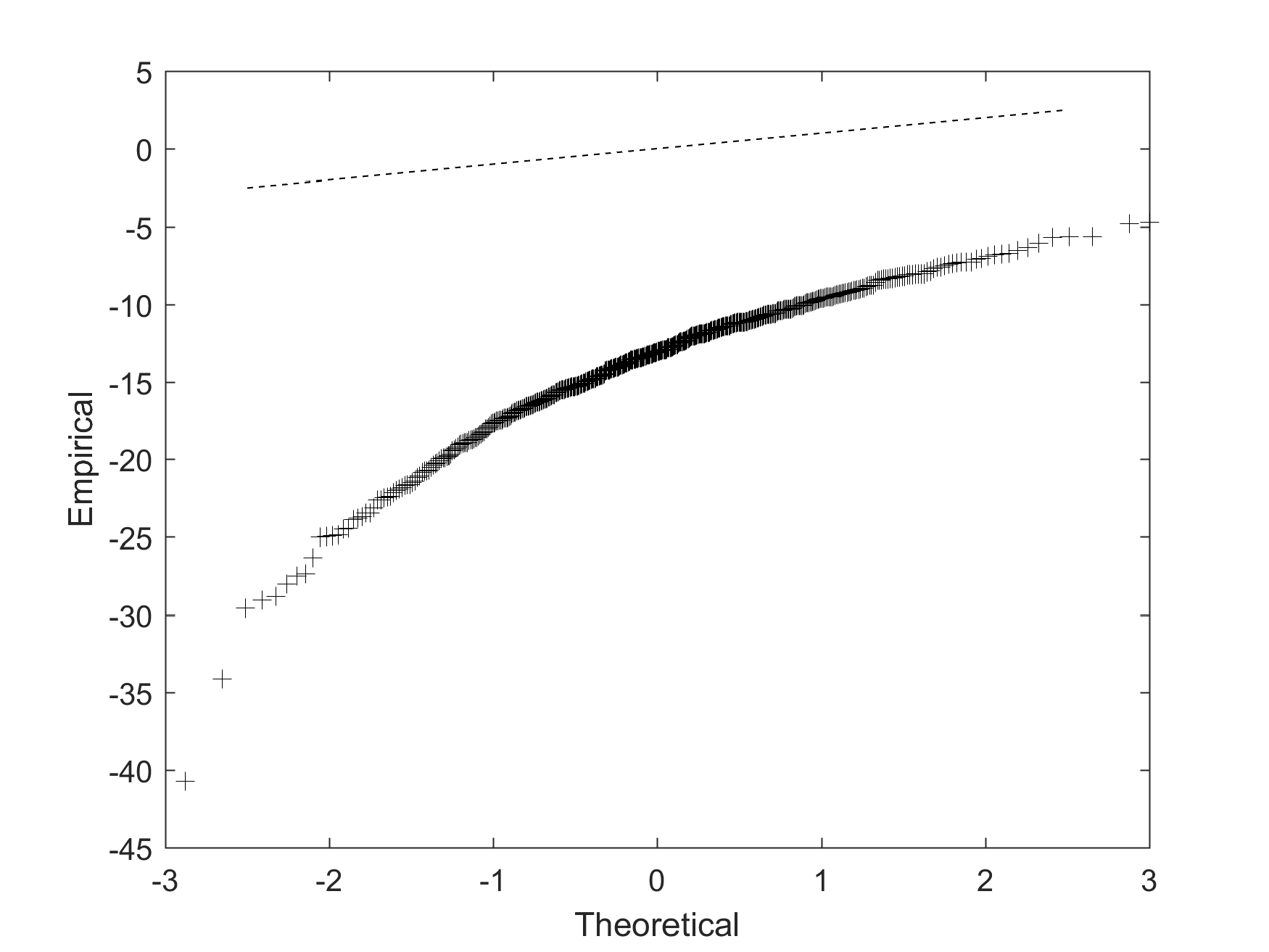} & \includegraphics[height=0.15\textheight,width=0.33\textwidth]{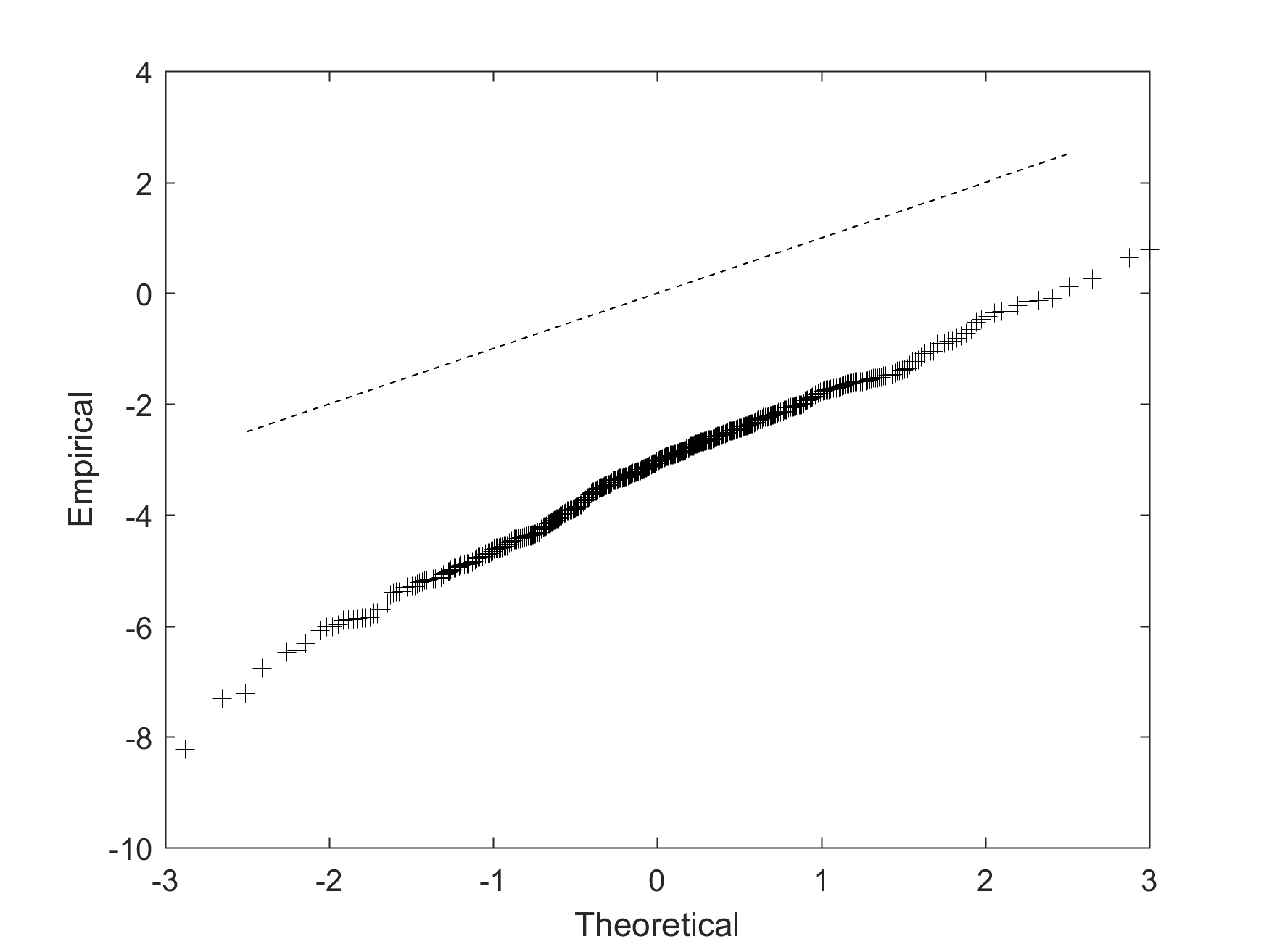} \\
		(g) & (h) & (i) \\
	\end{tabular}
	\caption{QQ-plots of the test statistic $\widehat{T}_i$ based on the two-stage IV model for several randomly selected variables to demonstrate the validity of its asymptotic distribution.  The panels in the first and second row correspond to selected variables whose true value are zero and the third row are variables that are not zero. For different columns, (a)(d)(g), (b)(e)(h) and (c)(f)(i) correspond to different $(n,p,q)$ values as $(200,100,100)$, $(400,200,200)$ and $(200,500,500)$.}
	\label{pic:single test for IV model}
\end{figure}

First, to show the validity of the asymptotic distribution of the proposed test statistic $\widehat{T}_i$ for single null hypothesis, 
we present in Figure \ref{pic:single test for IV model} the QQ-plots of the test statistics $\widehat{T}_{i}$ for several randomly selected covariates over in 500 replications, showing  that when using the correct two-stage IV model, the test statistic proposed follows a normal distribution under the null hypothesis  (panels (a)-(f)). However, for the covariates with non-zero distribution, the test statistic has a distribution that clearly deviates from the standard normal distribution (panels (g)-(i)).

\begin{figure}
	\centering
	\begin{tabular}{ccc}
		\includegraphics[height=0.15\textheight,width=0.33\textwidth]{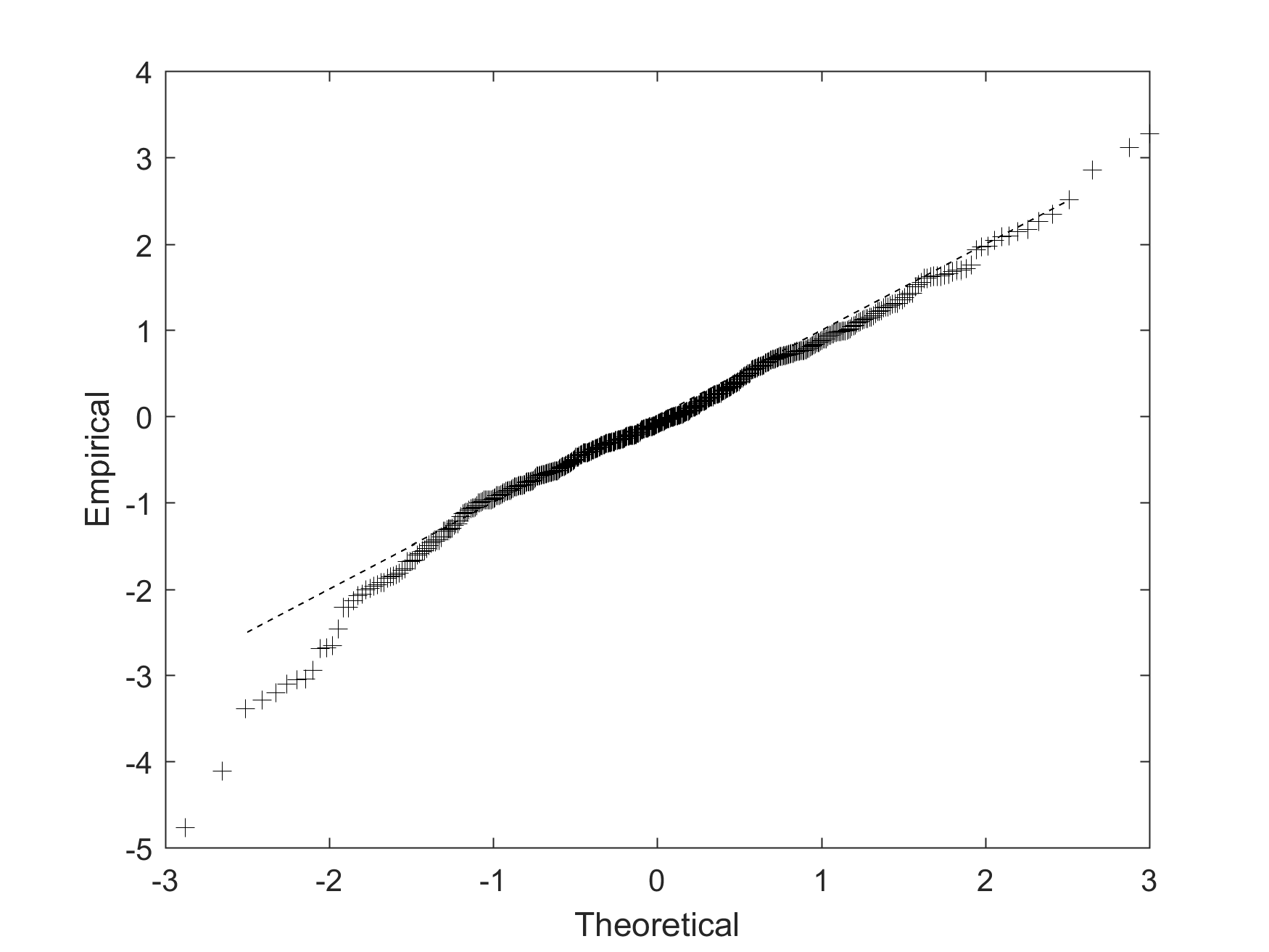} &	\includegraphics[height=0.15\textheight,width=0.33\textwidth]{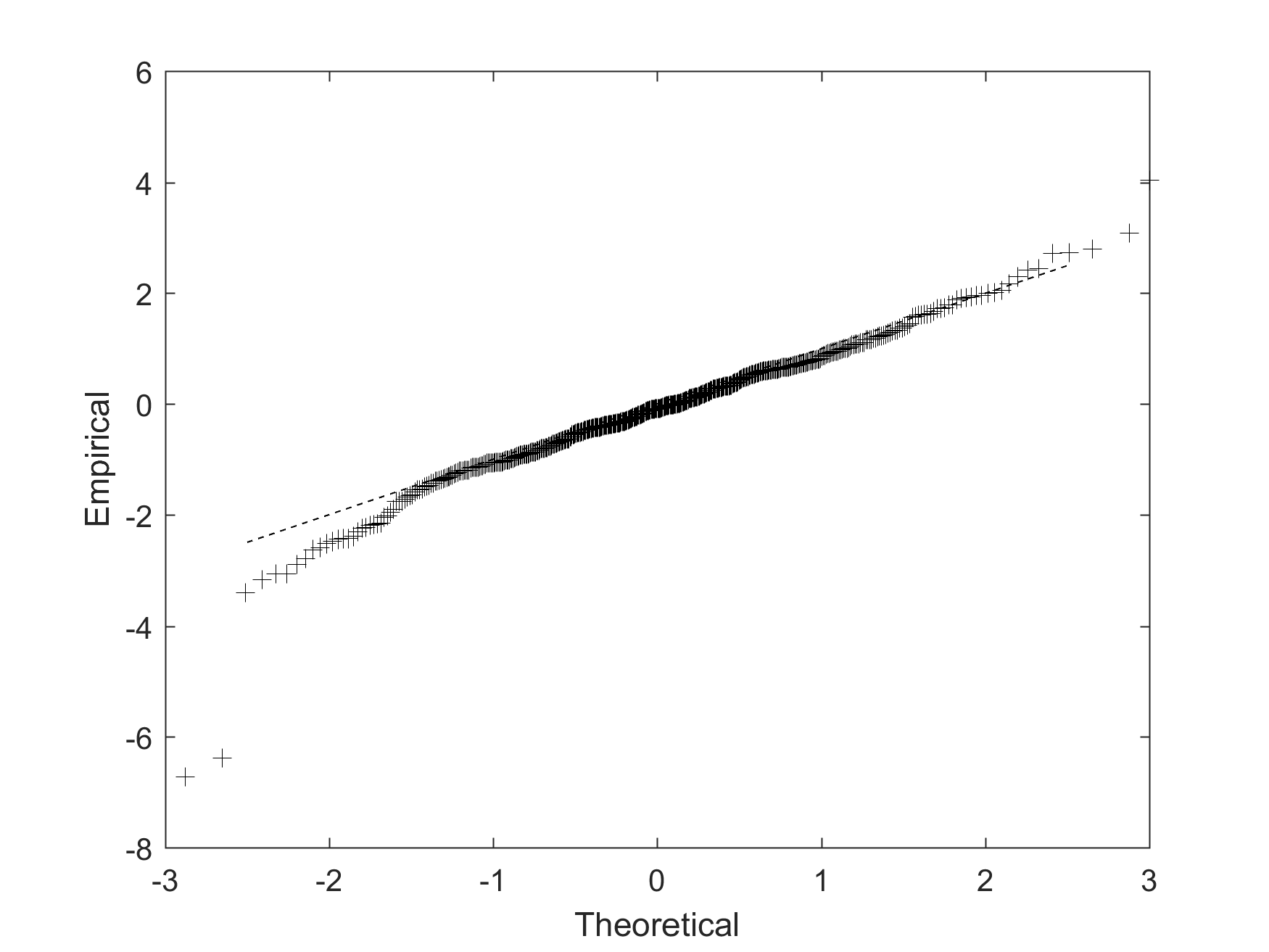} & \includegraphics[height=0.15\textheight,width=0.33\textwidth]{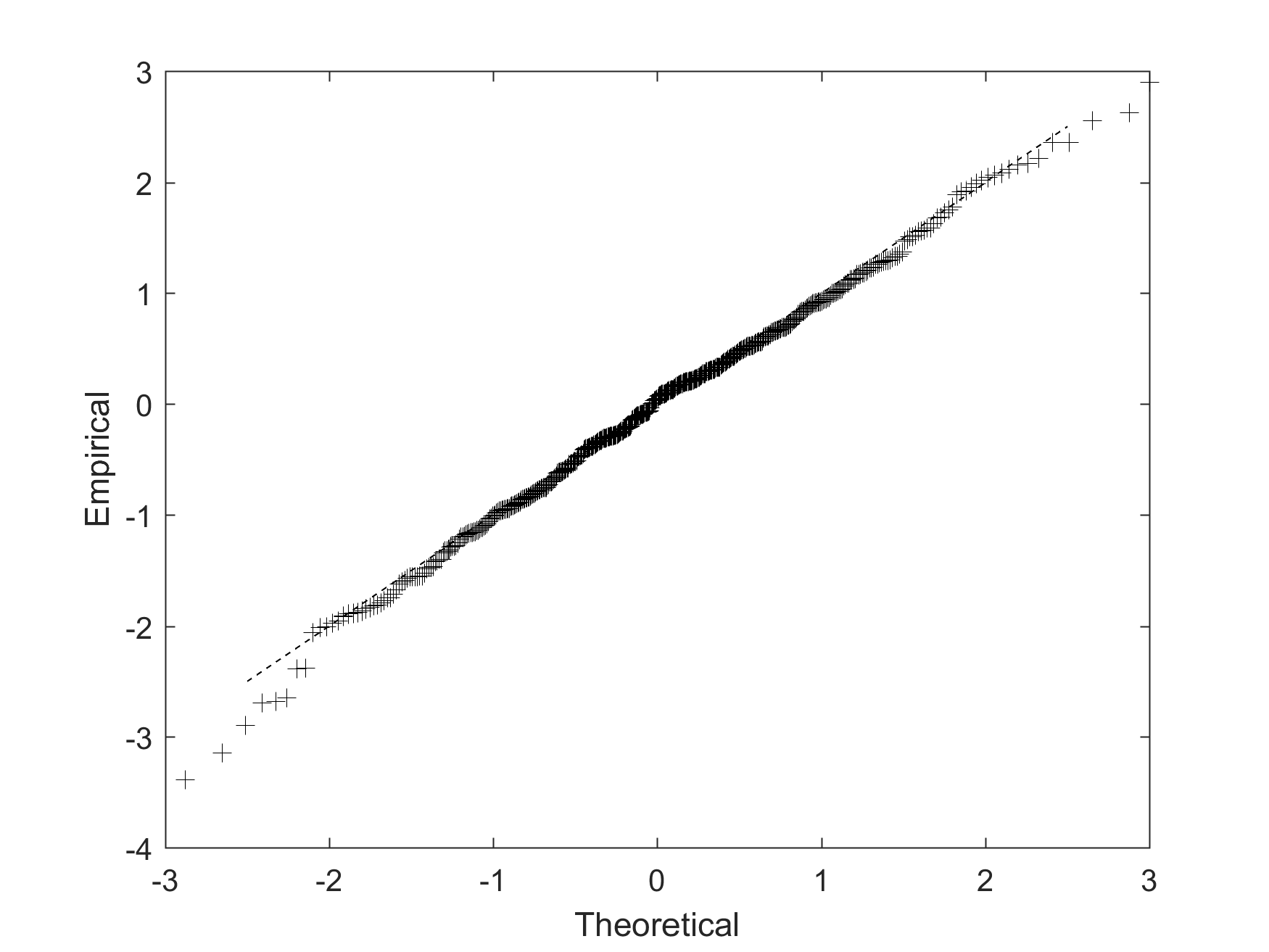} \\
		(a) & (b) & (c) \\
		\includegraphics[height=0.15\textheight,width=0.33\textwidth]{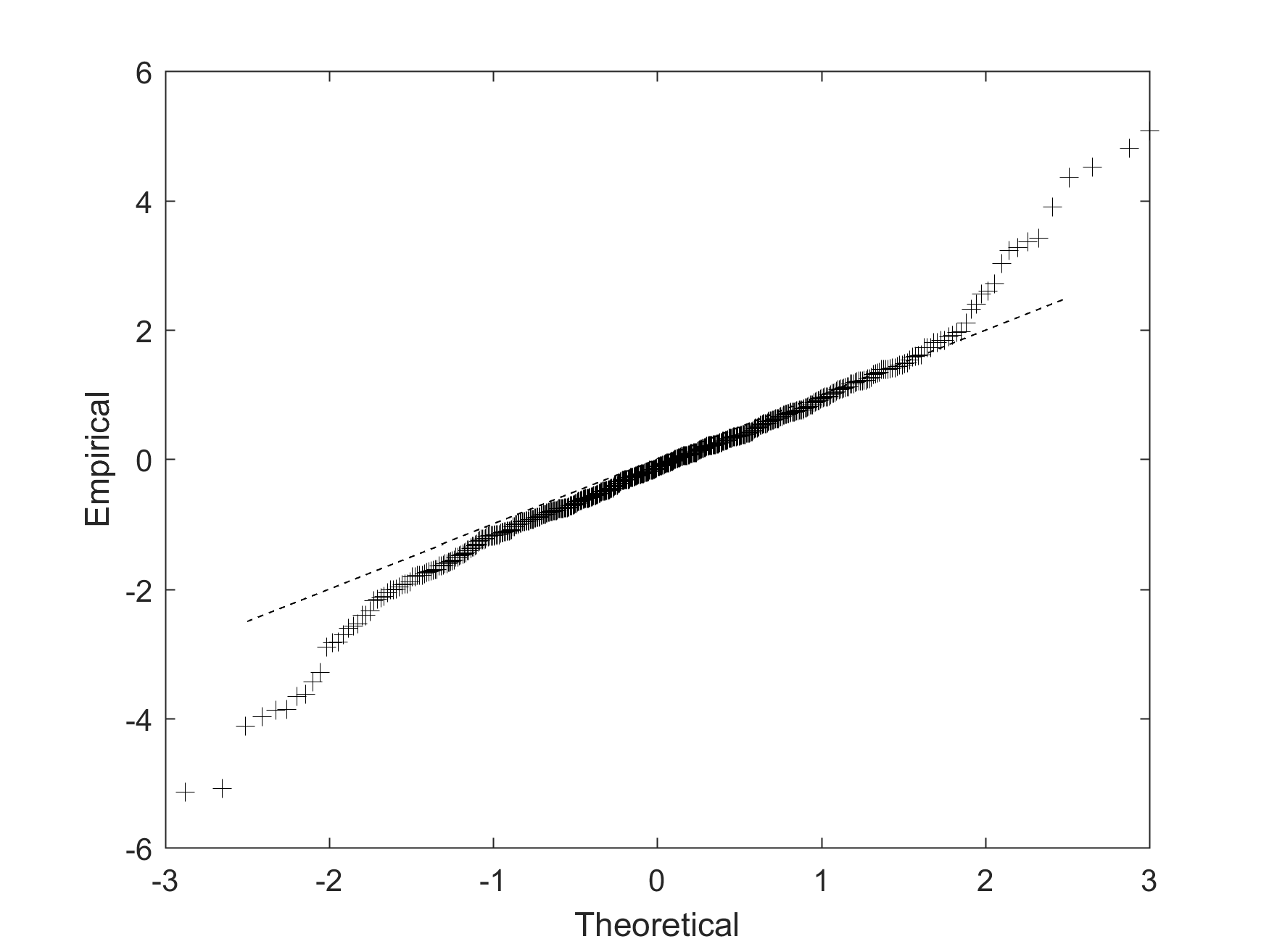} &	\includegraphics[height=0.15\textheight,width=0.33\textwidth]{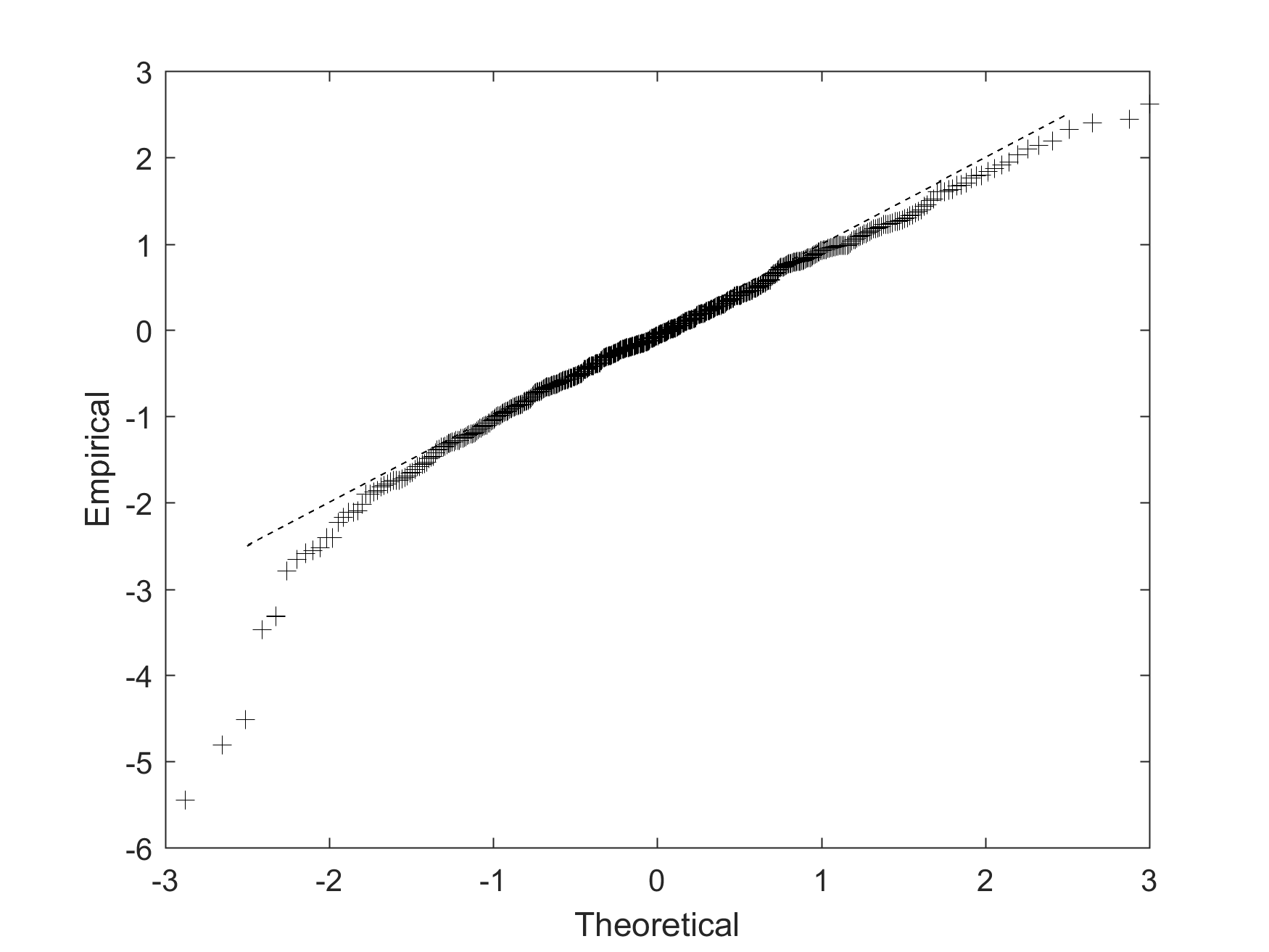} & \includegraphics[height=0.15\textheight,width=0.33\textwidth]{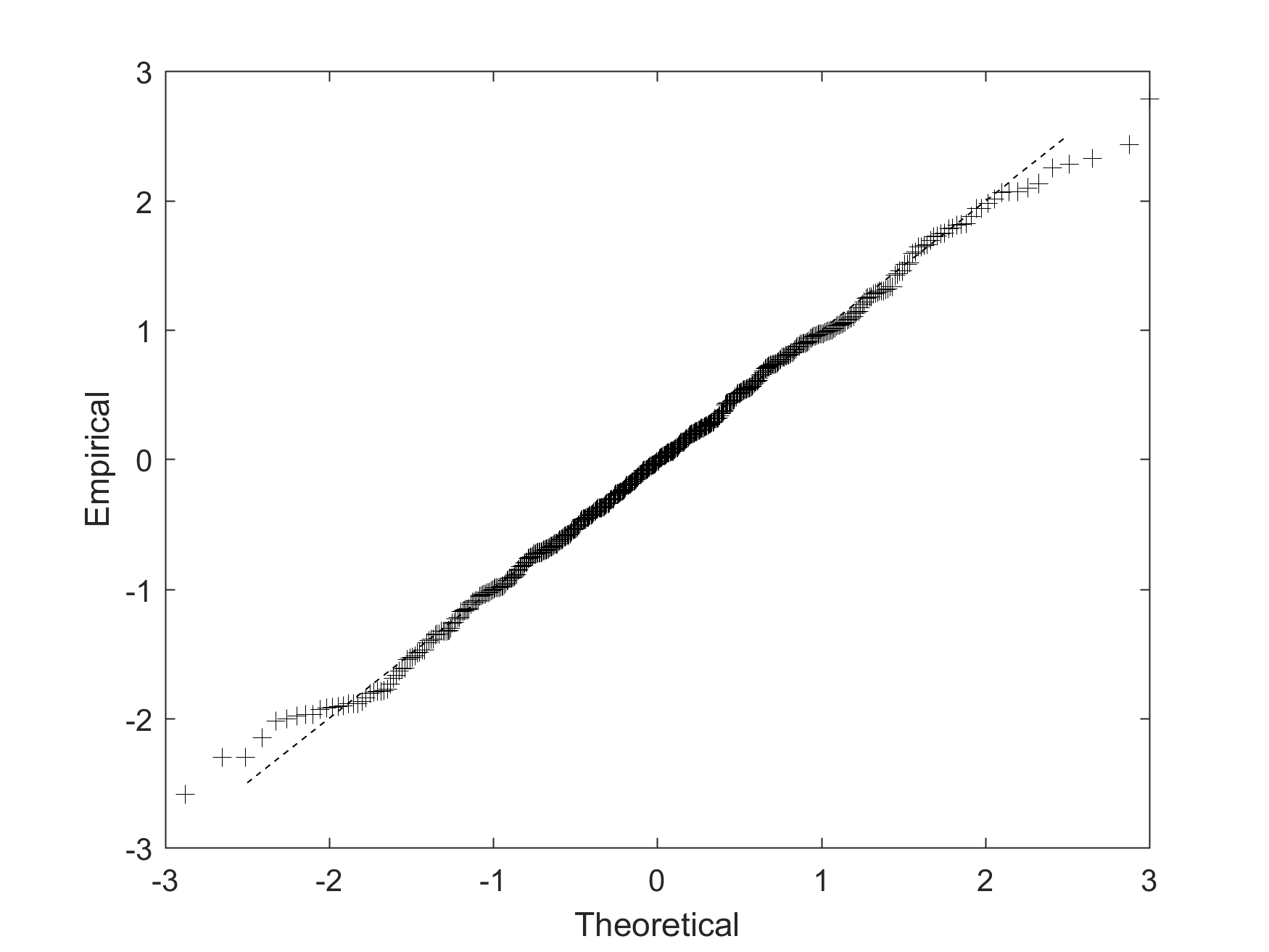} \\
		(d) & (e) & (f) \\
		\includegraphics[height=0.15\textheight,width=0.33\textwidth]{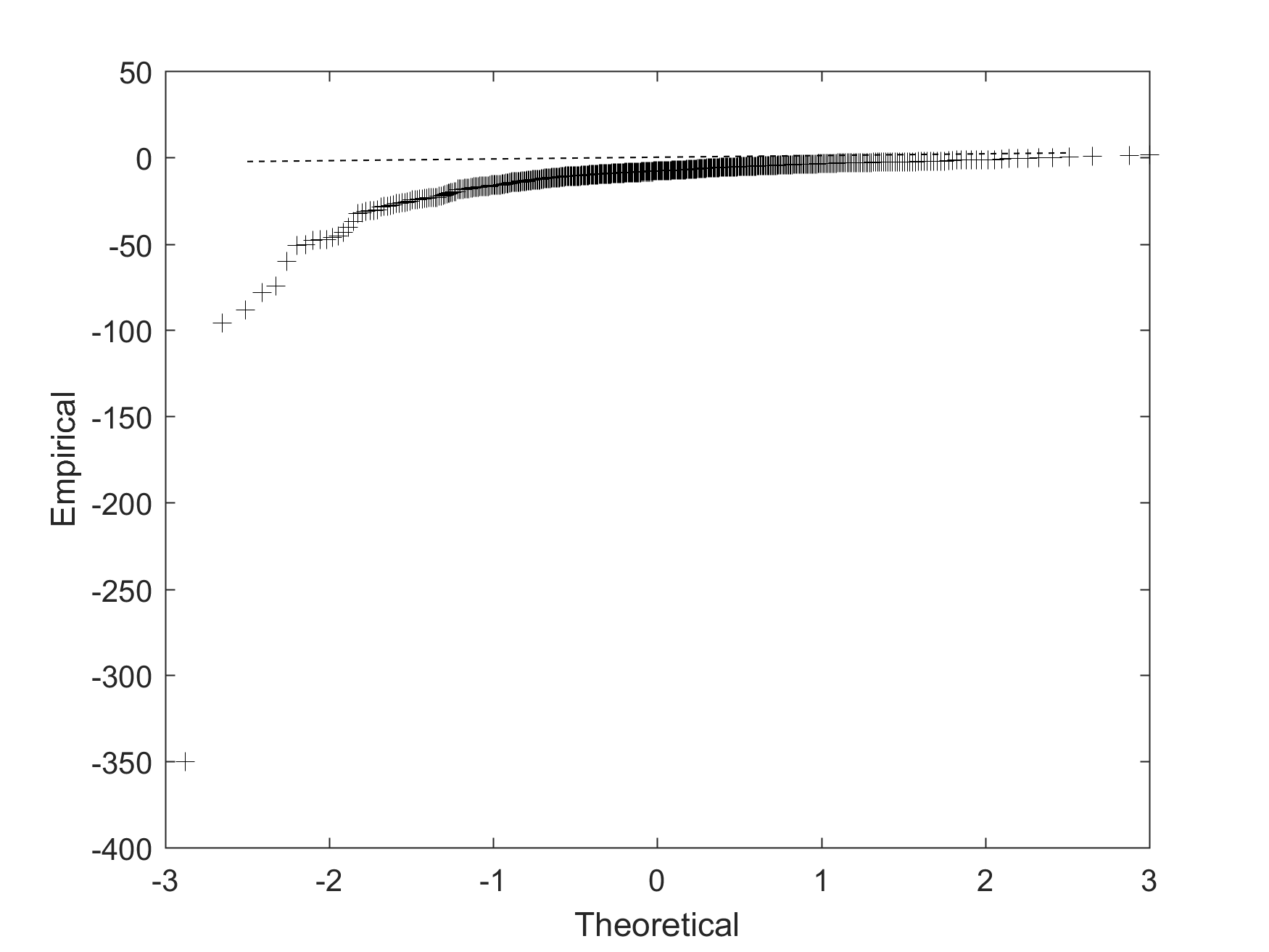} &	\includegraphics[height=0.15\textheight,width=0.33\textwidth]{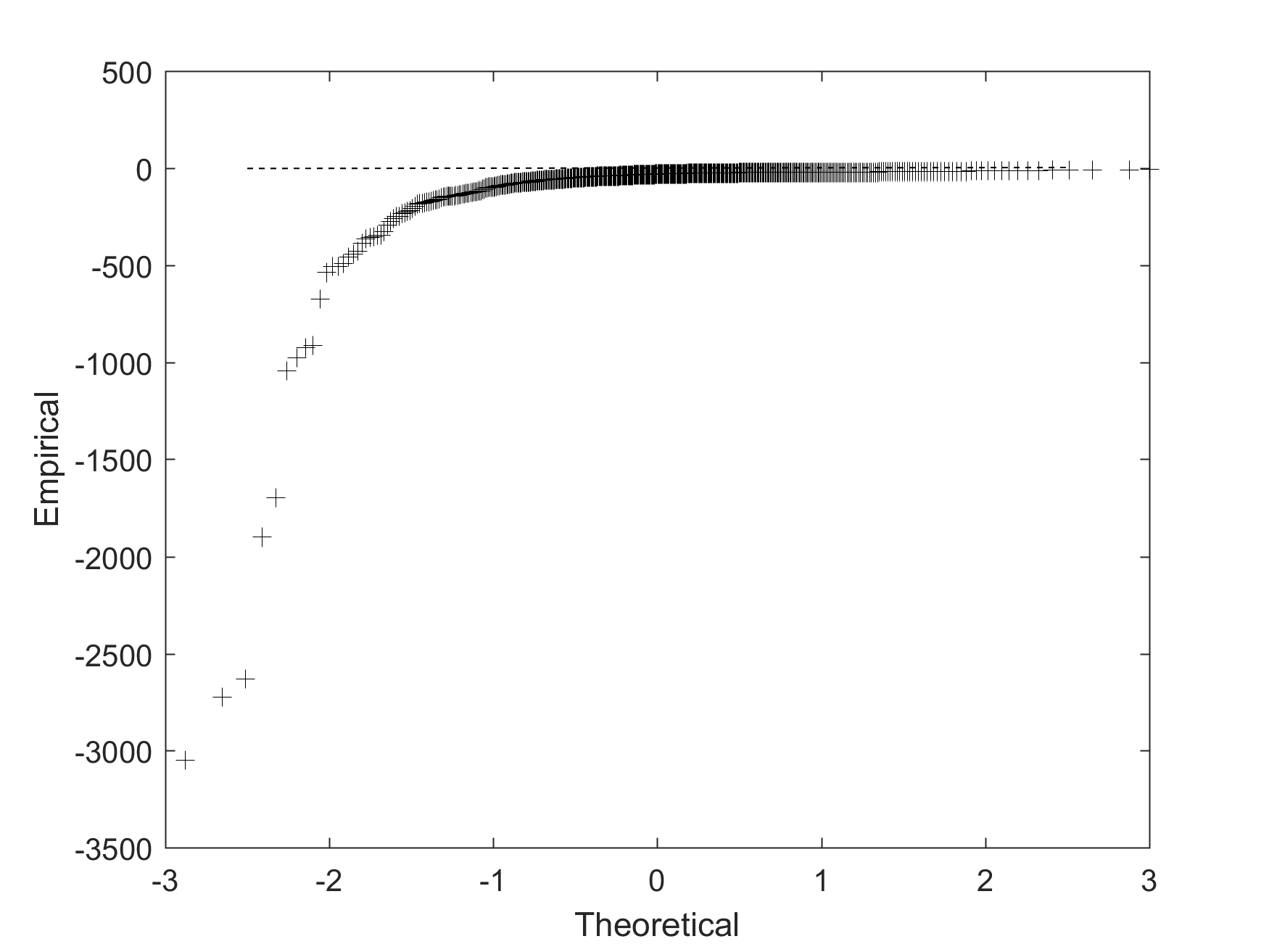} & \includegraphics[height=0.15\textheight,width=0.33\textwidth]{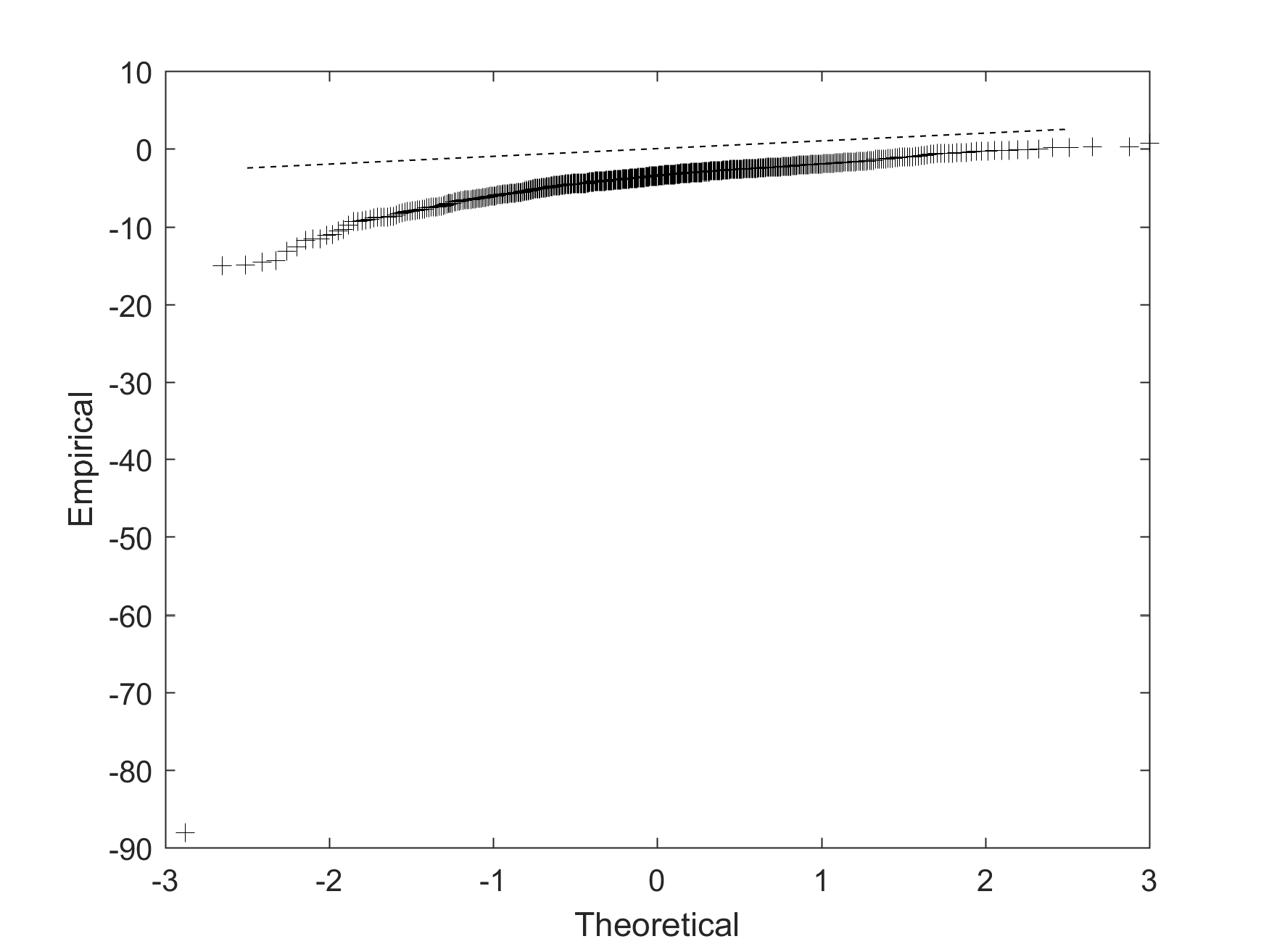} \\
		(g) & (h) & (i) \\
	\end{tabular}
	\caption{Selected QQ-plots of the test statistics $\widehat{T}_i$ developed for fitting naive high dimensional regression models.  The panels in the first and second row corresponds to selected variables whose true value are zero and the third row are variables that are not zero. For different columns, (a)(d)(g), (b)(e)(h) and (c)(f)(i) correspond to different $(n,p,q)$ values as $(200,100,100)$, $(400,200,200)$ and $200,500,500)$.}
	\label{pic:single test for wrong model}
\end{figure}

To demonstrate the importance of applying the IV model when the covariates and the error terms are dependent,  Figure \ref{pic:single test for wrong model}  shows the QQ-plots of the same set of variables as in the previous figure for the test statistic of \citet{liu2014hypothesis}. For the  variables with zero coefficients (panels (a)-(f)), the null distribution of the test statistic clearly deviates from the standard normal distribution for some  variables, indicating greater chance of identifying wrong variables.

\begin{figure}
	\centering
	\begin{tabular}{cc}
		\includegraphics[width=0.45\textwidth,height=0.2\textheight]{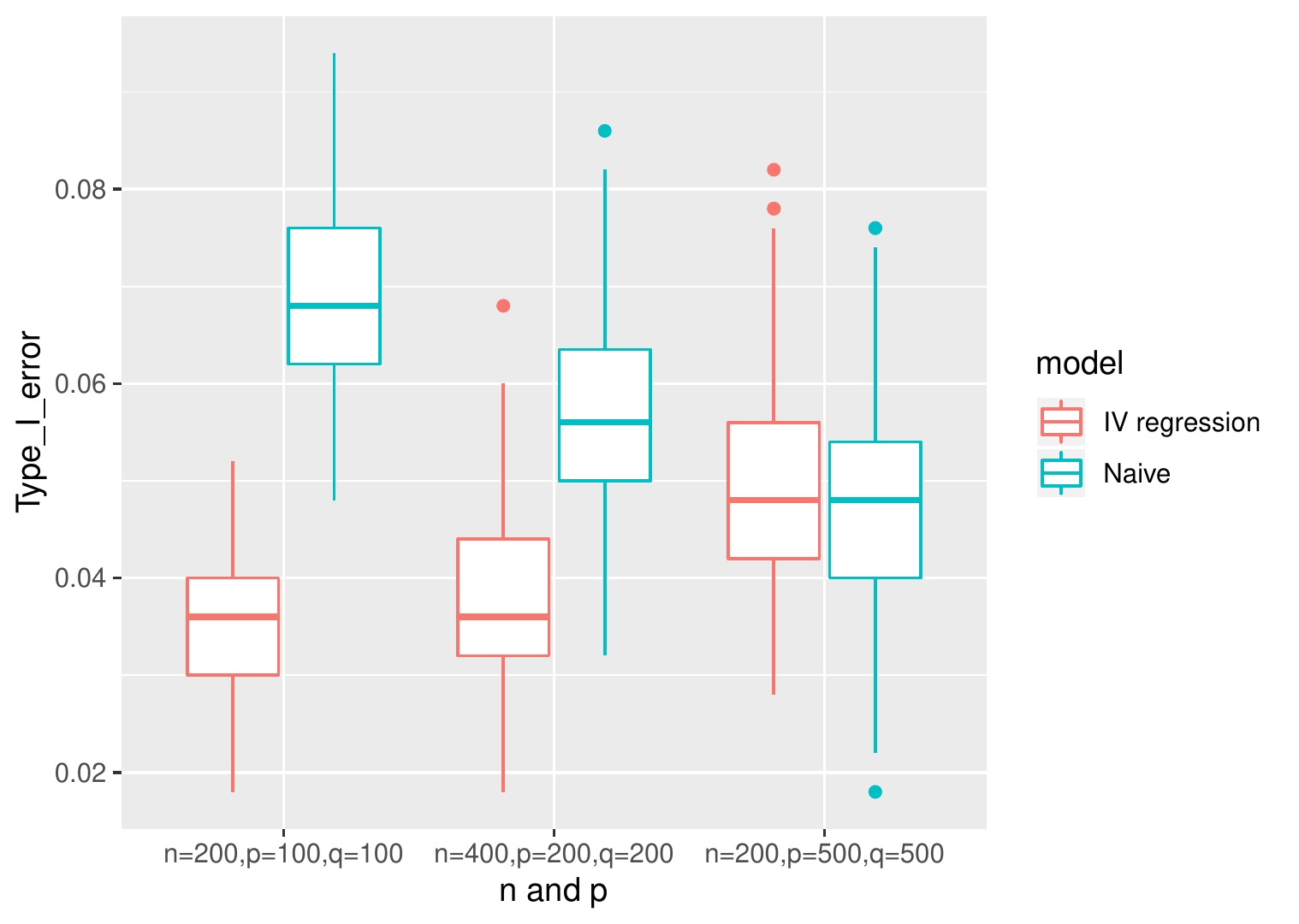} &	\includegraphics[width=0.45\textwidth,height=0.2\textheight]{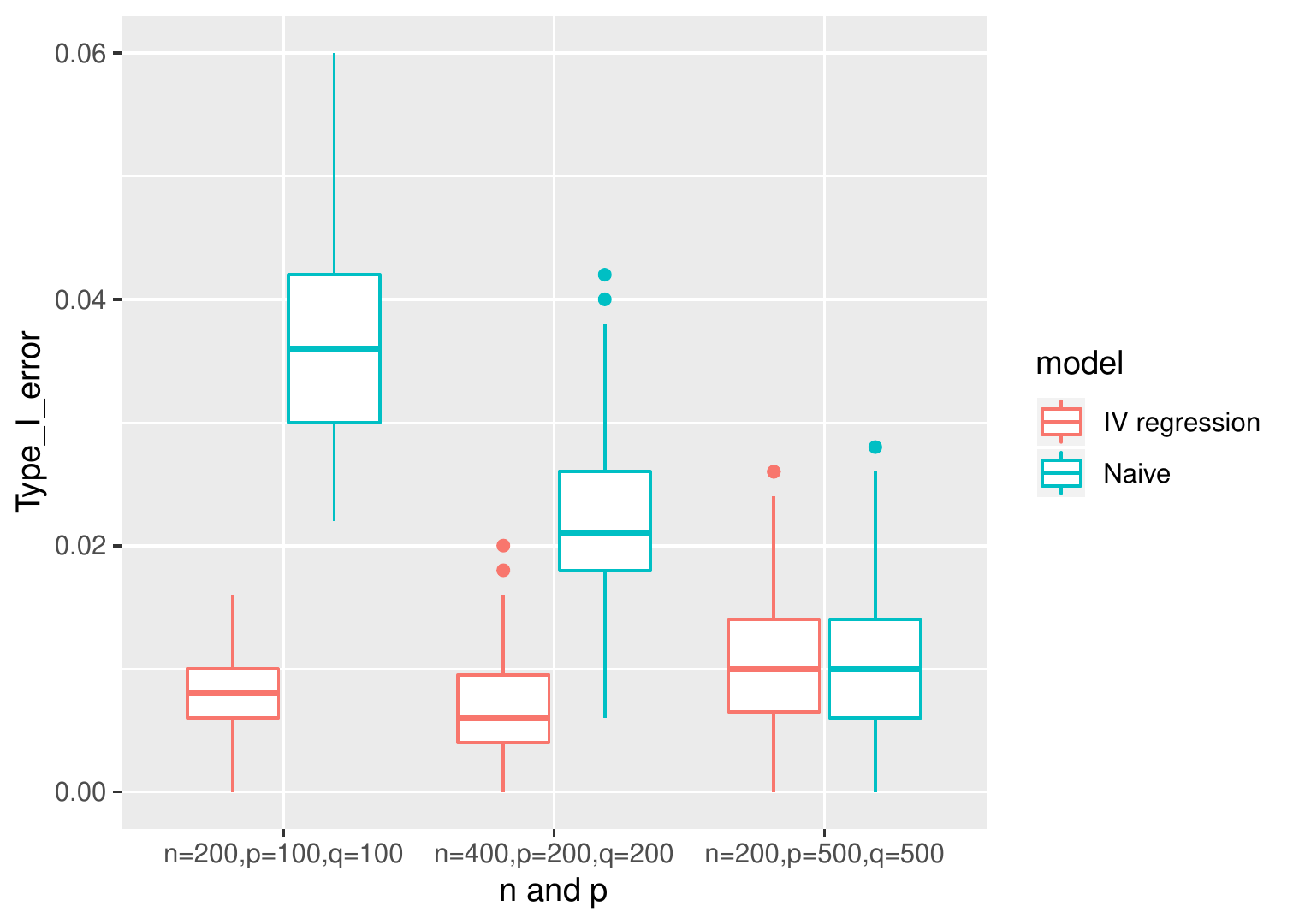} 
	\end{tabular}
	\caption{Box plots of the empirical type I errors  for single hypothesis testing based on IV regression and naive Lasso regression under different settings for $\alpha$-level of 0.05 (left) and 0.01 (right). }
	\label{plot: single test1}
\end{figure}

Figure  \ref{plot: single test1}  shows the box plots of the empirical type I errors for testing the single null hypothesis for the variables with zero coefficient based on IV models and the standard Lasso regression.  When the errors and covariates are correlated due to unobserved confounding,  the naive Lasso regression may fail to control the type I error for some null coefficients,  leading to inflated type I errors. This indicates that the naive method may falsely select some unrelated variables.  As a comparison,  the test based on the IV regression controls the type-I errors below the specified level.

\subsection{FDR Controlling for Multiple Testing}
To exam the performance of the proposed multiple testing procedure, the empirical FDR, defined  as 
\begin{align}
\text{eFDR} = \text{average(FDR)} \quad \text{where} \ \text{FDR} = \dfrac{\sum_{i \in \H_0} \bm{1}\left( |\widehat{T}_i | \geq \widehat{t}_0\right)}{\sum_{i=1}^{p}\bm{1}\left( |\widehat{T}_i | \geq\widehat{t}_0\right)\lor 1 },
\end{align}
is calculated. Similarly, the mean and standard deviation of the power defined as
\begin{align}
\text{power} = \dfrac{\sum_{i \in \H_1} \bm{1}\left( |\widehat{T}_i | \geq \widehat{t}_0\right)}{|\H_1|}.
\end{align}
The $\alpha$-level is chosen to be $\alpha = 0.05,0.1,0.2$. Table \ref{tab1} shows the empirical FDR for the proposed procedure using IV regression and the method of  \citet{liu2014hypothesis} using naive high dimensional regression models. The proposed multiple test procedure can indeed control the FDR at the correct level. In contrast, test based on naive high dimensional regression fails to control the FDR.

\begin{table}
	\centering
	\caption{Simulation results based on 500 replications. eFDR and power for multiple testing procedure based on IV regression and naive high dimensional linear regression for different combinations of $(n,p,q)$ and different $\alpha$ levels.   }
	\begin{tabular}{ccccc} 
		\hline 
		$n,p,q$ & $\alpha$-level & eFDR & power(sd)& eFDR (naive) \\[0.5 ex]
		\hline 
		\multirow{2}{5cm}{$(n,p,q)=(200,100,100)$} &	0.05  & 0.044	& 0.547	(0.15)   & 0.198   \\[0.5 ex]
		&	0.10  & 0.075  & 0.58 (0.15)  & 0.239   \\[0.5 ex]
		&	0.20  & 0.134  & 0.622 (0.15)  & 0.296   \\[0.5 ex]
		\hline 
		\multirow{2}{5cm}{$(n,p,q)=(400,200,200)$} &	0.05 &	0.026&	0.752 (0.13) &	0.153	 \\ [0.5 ex]
		&	0.10 	& 0.060 & 	0.781 (0.12) &	0.197\\ [0.5 ex]
		&	0.20	 & 0.124 & 	0.814 (0.12) &	0.268\\ [0.5 ex]
		\hline 
		\multirow{2}{5cm}{$(n,p,q)=(200,500,500)$} &	0.05 &	0.074&	0.390 (0.12) &	0.055	 \\ [0.5 ex]
		&	0.10 	& 0.129 & 	0.427 (0.13) &	0.103\\ [0.5 ex]
		&	0.20	 & 0.224 & 	0.472 (0.14) &	0.197 \\ [0.5 ex]
		\hline   
	\end{tabular}
	\label{tab1} 
\end{table}

We similarly evaluated the procedure for  controlling  the number of falsely discovered variables. The empirical  FDV is defined as
\begin{align*}
\text{eFDV} = \text{average(FDV)} \quad \text{where} \ \text{FDV} = \sum_{i \in \H_0} \bm{1}\left( |\widehat{T}_i | \geq \widehat{t}_{FDV}\right),
\end{align*}
and its power is given by
\begin{align*}
\text{power} = \sum_{i \in \H_1} \bm{1}\left( |\widehat{T}_i | \geq \widehat{t}_{FDV}\right).
\end{align*}
We consider the $k$-level of 2,3 and 4.  Table  \ref{tab2} shows that the proposed procedure  also controls the FDV at the specified  level. However, naive test that  ignoring the covariate-error dependence can result in failing to control the FDV.

\begin{table}
	\centering
	\caption{Simulation results based on 500 replications. eFDV and power for multiple testing procedures based on IV regression and naive high dimensional linear regression for different  combinations of $(n,p,q)$ and and different $k$  levels.}
	\begin{tabular}{ccccc} 
		\hline 
		$n,p,q$ & $k$-level & eFDV& power (sd)& eFDV (naive) \\[0.5 ex]
		\hline 
		\multirow{2}{5cm}{$(n,p,q)=(200,100,100)$} &	2 &	1.35 &	6.35 (1.5) &	4.11	 \\ [0.5 ex]
		&	3	&1.94 & 	6.57 (1.4) &	4.87\\ [0.5 ex]
		&	4 & 2.49 & 	6.71 (1.4) &	5.55\\ [0.5 ex]
		\hline 
		\multirow{2}{5cm}{$(n,p,q)=(400,200,200)$} &	2  & 1.27& 8.16	(1.1)   & 4.18   \\[0.5 ex]
		&	3 & 1.94  & 8.31 (1.1)  & 5.13   \\[0.5 ex]
		&	4  & 2.59  & 8.42 (1.1)  & 5.96   \\[0.5 ex]
		\hline 
		\multirow{2}{5cm}{$(n,p,q)=(200,500,500)$} &2 &	2.21&	4.93 (1.3) &	2.04	 \\ [0.5 ex]
		&	3	& 3.19 & 	5.17 (1.4) &	3.01 \\ [0.5 ex]
		& 4	 & 4.13 & 	5.39 (1.4) &	3.98 \\ [0.5 ex]
		\hline   
	\end{tabular}
	\label{tab2} 
\end{table}

It is worth noting that  for $p=500$, the performance of our proposed method is very similar to the naive test. The reason is that by our construction of the covariance matrix of the error terms, the dependence between covariates and errors  becomes very week for large $p$, in which case  the two methods  are expected to perform similarly. 

\section{Application to a Yeast Data Set}\label{sec: application}
We demonstrate our method using a data set  collected on 102 yeast segregants created by  crossing  of two genetically diverse strains \citep{yeast2}. The data set includes the growth yields of each segregant grown in the presence of different chemicals or small molecule drugs \citep{yeastdrug}. 
These segregants have different genotypes represented by 585 markers  after removing the markers that are in almost complete linkage disequilibrium. The genotype differences in these strains contribute to rich phenotypic diversity in the segregants. In addition,  6189 yeast genes were profiled in rich media and in the absence of any chemical or drug using expression arrays  \citep{yeast2}.  Using the same data preprocessing steps as \cite{yeast1}, we compiled a list of candidate gene expression features based on their potential regulatory effects, including transcription factors, signaling molecules, chromatin factors and RNA factors and  genes involved in vacuolar transport, endosome, endosome transport and vesicle-mediated transport.  We further filtered  out the genes with $s.d\le 0.2$ in expression level, resulting a total of  813 genes in our analysis.

\begin{figure}
	\centering
	\begin{tabular}{ccc}
		\includegraphics[height=0.2\textheight,width=0.45\textwidth]{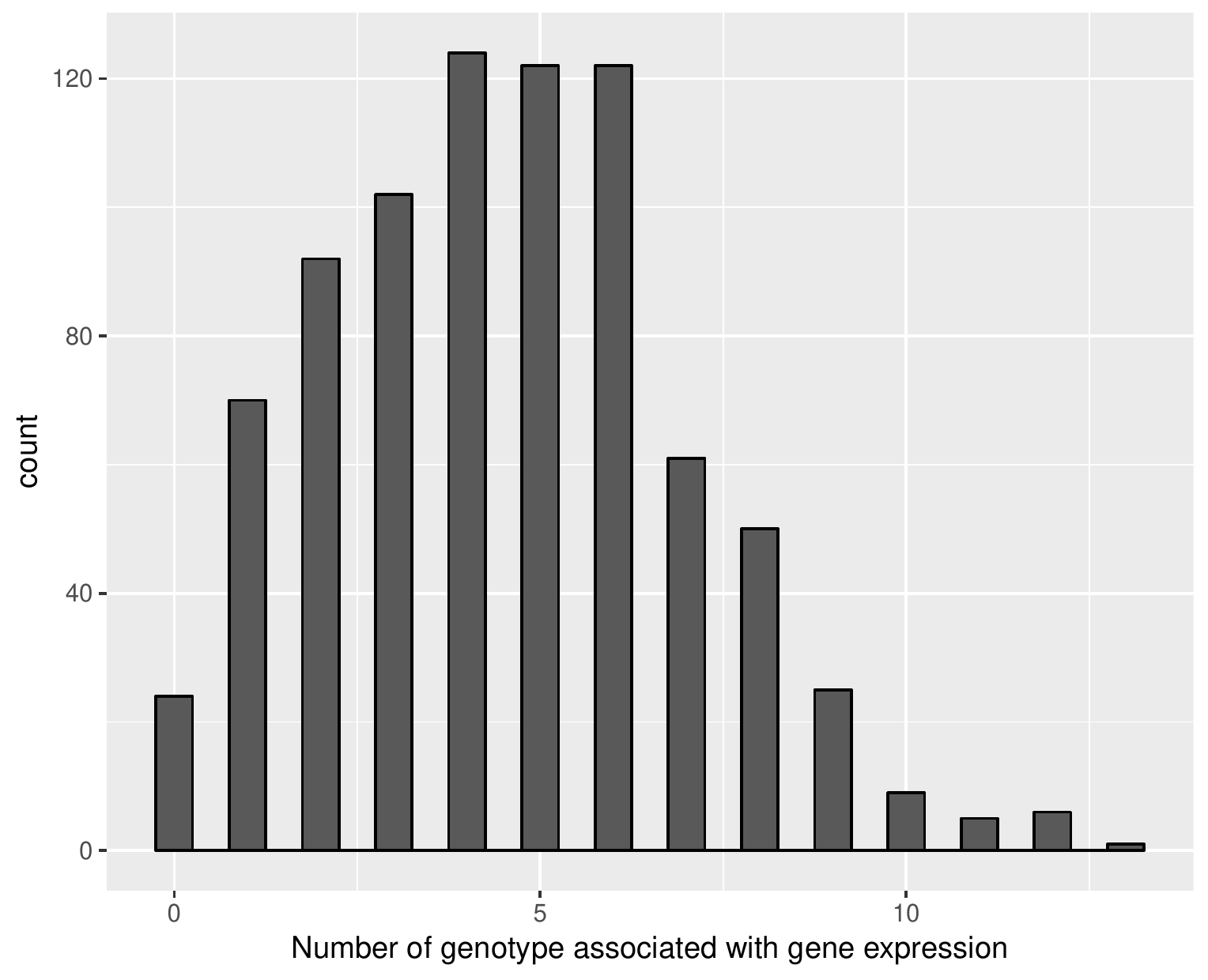} &	\includegraphics[height=0.2\textheight,width=0.45\textwidth]{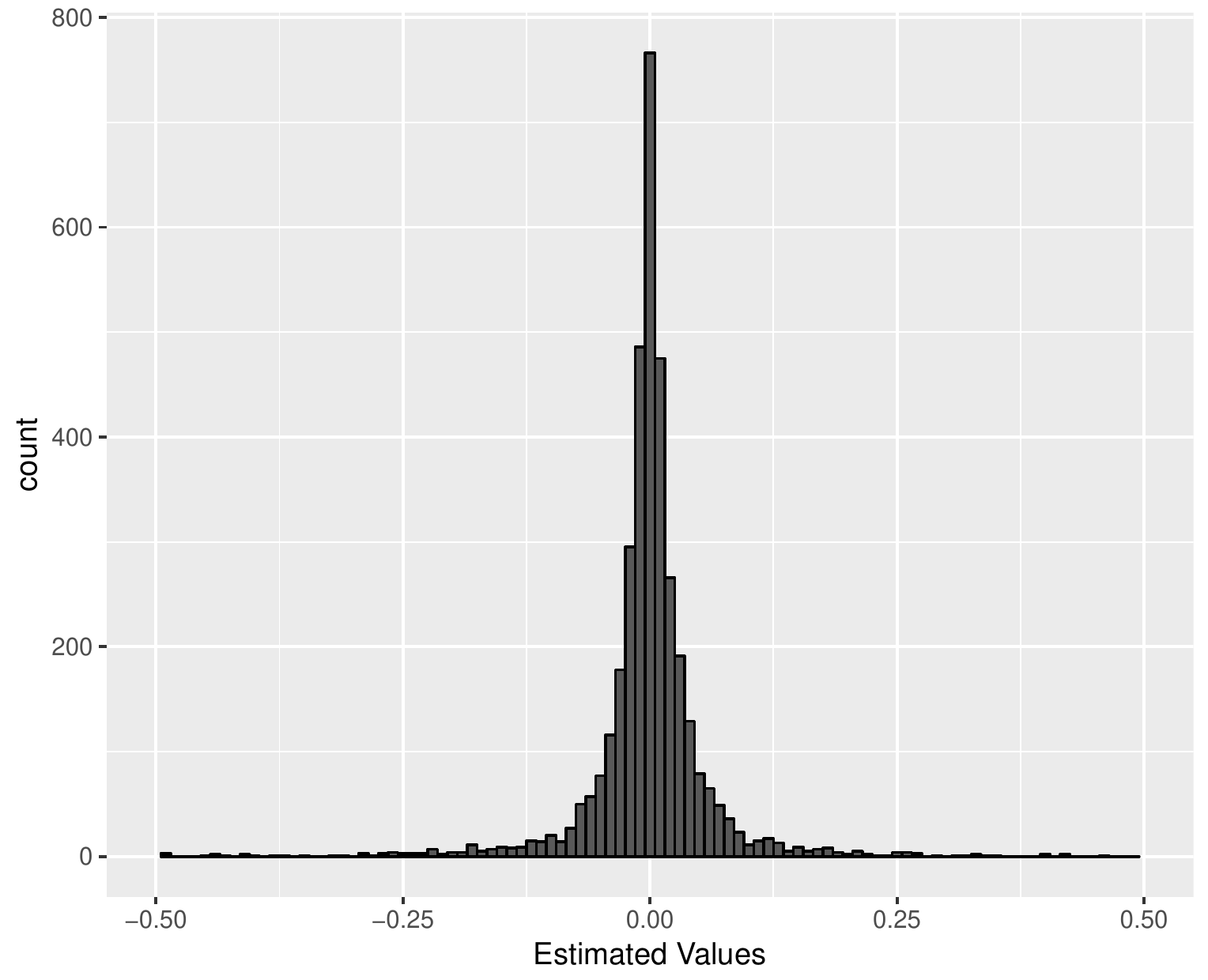} & 
	\end{tabular}
	\caption{Analysis of yeast eQTL data sets, showing the  histogram of the number of genotypes associated with each gene expression (left plot) and  the histogram of the estimated regression coefficients in the first stage ($\widehat{\bGamma}$) based on Lasso regressions (right plot).}
	\label{pic:IV and estimated value}
\end{figure}

We are interested in identifying the genes whose expression levels are associated with  yeast growth yield after being treated with hydrogen peroxide by fitting the proposed two-stage sparse IV model. 
Figure \ref{pic:IV and estimated value} shows the histogram of the number of SNPs  selected for each gene expression and the histogram of the estimated regression coefficients ($\bGamma_0$) from Lasso.  These results show that genetic variants are strongly associated with gene expressions and therefore can be used as instrument variables for gene expressions.

Using these selected genotypes as the instrumental variables for each of the gene expressions, we obtained the fitted expression values and applied Lasso with these fitted expressions as predictors and yeast growth yield as the response. For each gene $j$, we tested the null of $\beta_j=0$  and obtained its $p$-value. The 15 significant genes at a nominal $p<0.05$ are presented in Table \ref{yeast.res1}. At FDR$<0.10$, three genes were selected.  These genes are related with resistance to chemicals, competitive fitness and cell growth, partially explaining their association with the yeast growth in the presence of hydrogen peroxide. 
For example, among the genes with negative coefficient, over-expression indicates decreased yeast growth. RRM3 gene is involved in DNA replication, and over-expression of the gene leads to abnormal 
budding and decreased resistance to chemicals.   Over-expression of POP5 and FUN26 genes causes decreased vegetative growth rate of yeast  (\url
{https://www.yeastgenome.org}).

The three selected genes using FDR$<0.10$ all had  positive coefficients, indicating 
over-expression of these genes led to increased yeast growth in the presence of  hydrogen peroxide.
Among these, BDP1 is a general activator of RNA polymerase III transcription and is required for transcription from all three types of polymerase III promoters \citep{BDP1}, and over-expression of this gene is expected to increase the yeast viability and growth.  PET494 is a mitochondrial translational activator specific for mitochondrial mRNA  encoding cytochrome c oxidase subunit III (coxIII) 
\citep{Pet494}.   Finally, null mutant of ARG4 gene shows decreased resistance to chemicals (\url{https://www.yeastgenome.org}) and therefore segregants with higher expression of this gene are expected to have increased resistance to chemicals and increased growth yield. 

\begin{table}
	\centering
	\caption{Results from analysis of yeast growth yield data. Table shows the selected genes using single test statistics ($p<0.05$) and multiple testing procedure with FDR$< 0.10$ and FDV$<2$(marked by $*$). The gene names and estimated regression coefficients and refitted values are listed.}
	\begin{tabular}{llll}
		\hline
		Gene id  & Gene name & $\hat{\bbeta}$  & Refitted $\hat{\bbeta}$ \\
		\hline
		\multicolumn{4}{c}{Negative  coefficient}\\	
		YHR031C    & RRM3  & -3.82 & -5.00 \\
		YAL033W    &   POP5 & -0.22 & -0.69 \\
		YLR275W   &  SMD2 & -0.20 & -0.31 \\
		YNL236W    & SIN4  & -4.67 & -5.63 \\
		YNL138W    &   SRV2 & -0.63 & -1.68 \\
		YNL146W  &  YNL146W & -0.24 & -0.12 \\
		YAR035W   &   YAT1 & -1.74 & -2.79 \\
		YAL022C    & FUN26 & -2.89 & -4.79 \\
		YHL018W   &  YHL018W & -0.79 & -2.29 \\
		\multicolumn{4}{c}{Positive coefficient}\\
		YNL331C    &  AAD14 & 0.07  & 0.17 \\
		YHR014W   &  SPO13 & 0.47  & 2.20 \\
		YHR018C$^{*}$    &  ARG4 & 0.22  & 0.34 \\
		YHR097C   & YHR097C & 0.06  & 0.15 \\
		YNL039W$^{*}$     &  BDP1 & 1.82  & 3.96 \\
		YNR045W$^{*}$   & PET494 & 0.70  & 0.86 \\
		\hline
	\end{tabular}
	\label{yeast.res1}
\end{table}

As a comparison,  we also applied  Lasso regression with 813 gene expressions as the predictors without using the genotype data.  The same statistical test was applied to each of the genes.  At a nominal $p$-value of 0.05,  34 genes were selected by Lasso. However, no gene was selected after adjusting for multiple comparisons with  FDR$<0.10$. This suggests that by effectively using the genotype data, we were able to identify biologically meaningful genes that are associated with yeast growth in the presence of hydrogen peroxide.

\begin{figure}
	\centering
	\begin{tabular}{ccc}
		\includegraphics[height=0.2\textheight,width=0.3\textwidth]{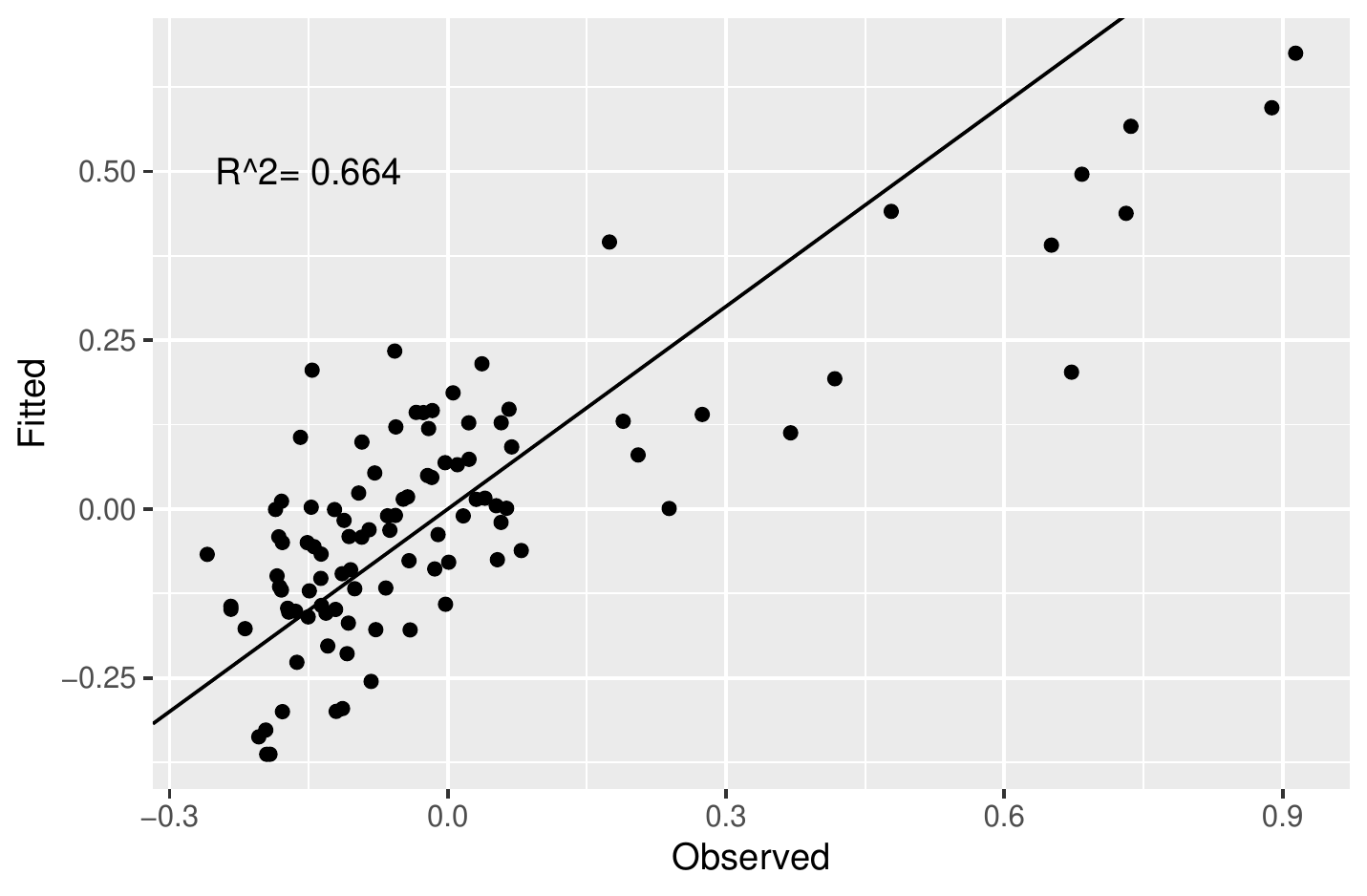} &	\includegraphics[height=0.2\textheight,width=0.3\textwidth]{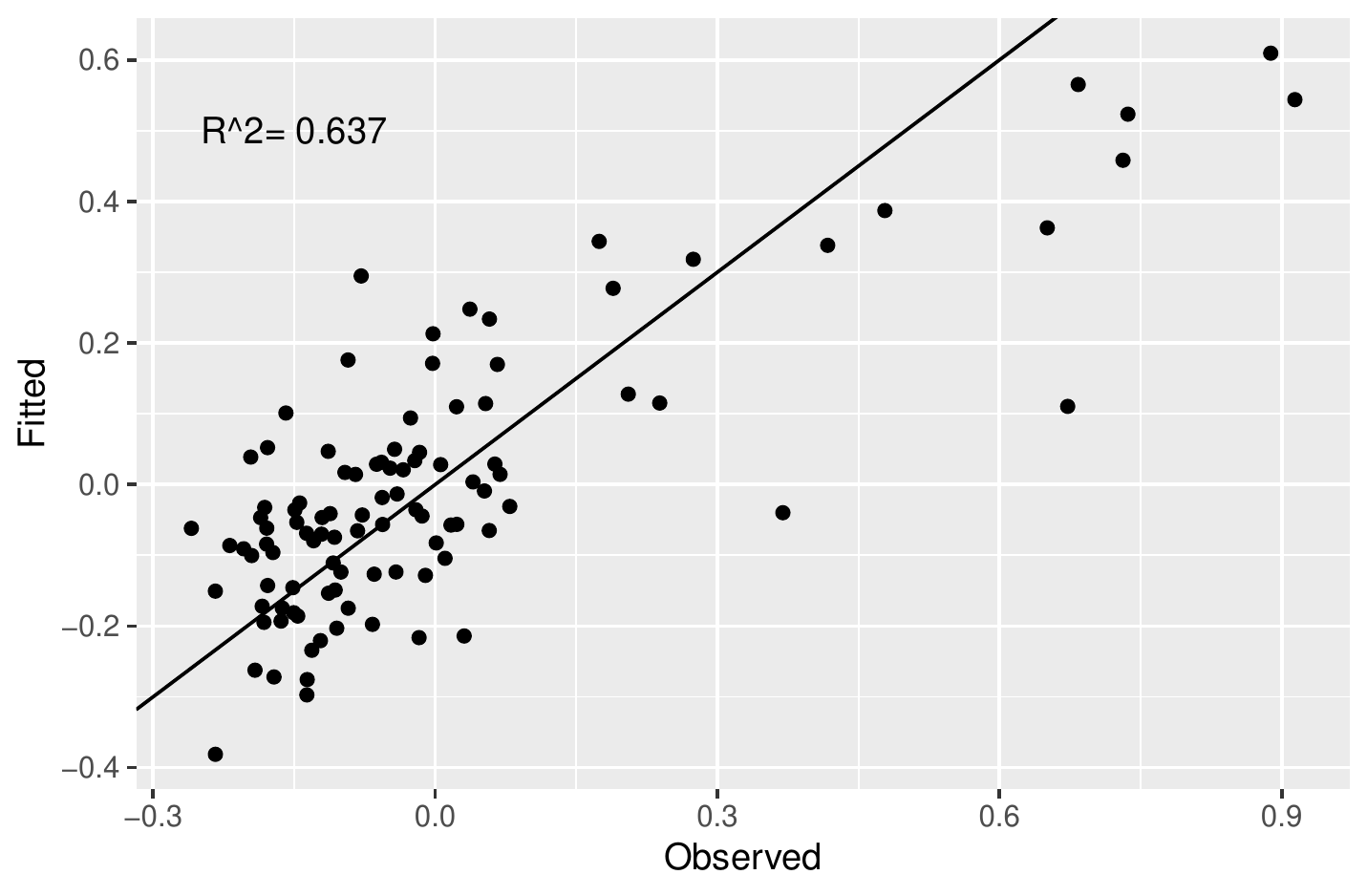} & \includegraphics[height=0.2\textheight,width=0.3\textwidth]{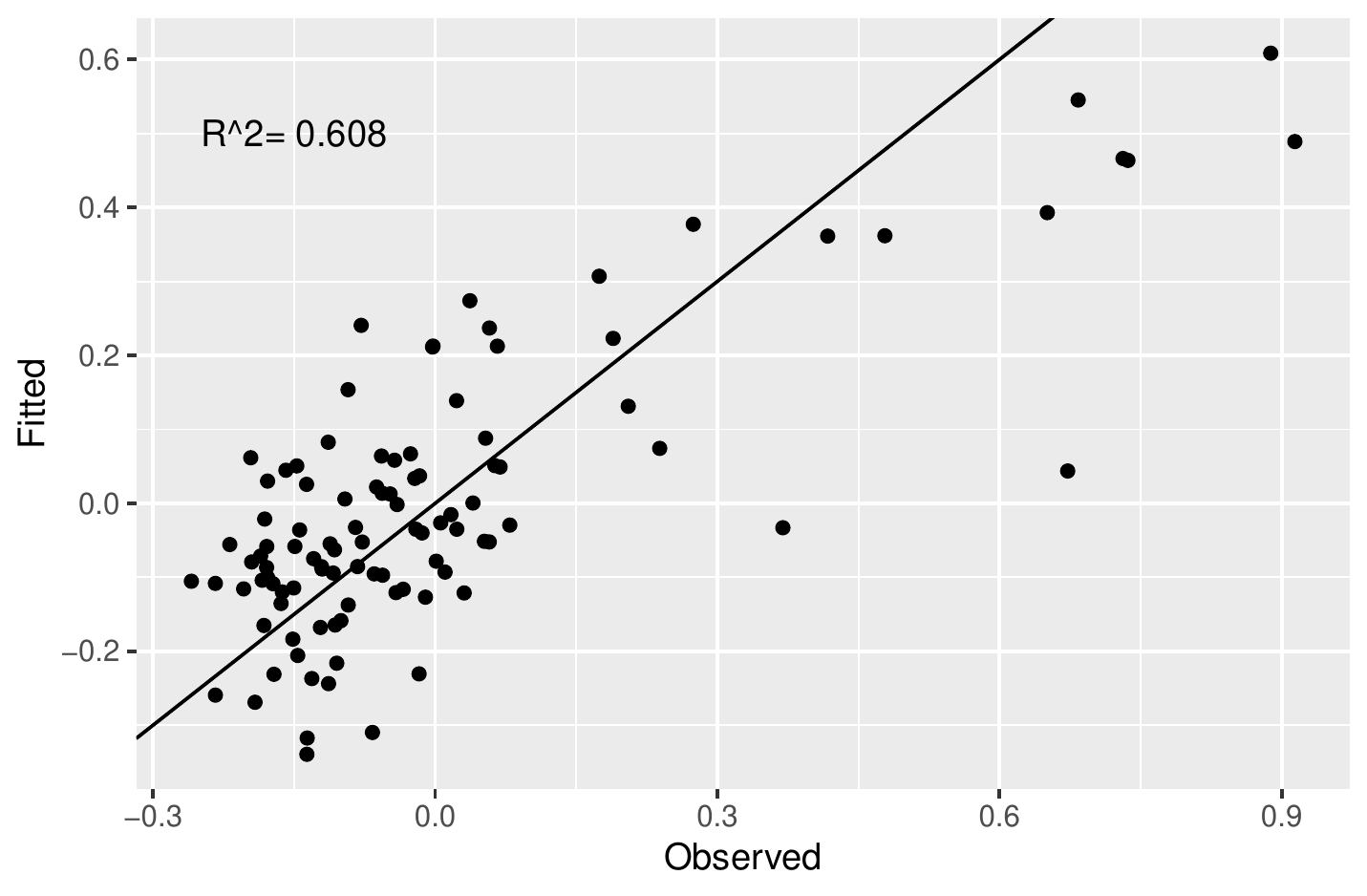} \\
		(a) & (b) &(c)
	\end{tabular}
	\caption{Scatter-plots of the fitted versus the observed yeast growth yield. (a):  refitted  model using the estimated  expression levels of the 15 genes selected by our proposed method;  (b):  refitted model using expression levels of 34 genes selected using naive test;  (c): the refitted model using expression levels of 34 genes selected based on Lasso.}
	\label{pic: refitted}
\end{figure}

We further compared the model fits by calculating  the $R^2$ statistics in three different scenarios. The first scenario is to use the 15 genes selected using our proposed multiple testing method and refit a linear model with the estimated $\widehat{\bX}$. The second scenario is use the 34 genes identified by  naive test  and refit a linear model using the original $\bX$. The last scenario  is use the genes selected by Lasso using  $\bX$ and refit a linear model with the original $\bX$. Figure \ref{pic: refitted}  shows that our method provides the highest $R^2$ value among the three, with a value of $0.664$, indicating that  using refitted $\bX$ can lead to  better fit of the data. 

\section{Discussion} \label{sec: discussion}
We have  developed methods for exploring the association between gene expression and phenotype in the framework IV regression when there are possible unmeasured confounders. Here the   genetic variants are used as possible instrumental variables.  We   have constructed a test statistic using the idea of inverse regression and derived  its asymptotic null  distribution. We have further  developed  a multiple testing procedure for the high-dimensional two stage least square methods and provided the rejection region of multiple testing that controls the false discovery rate or number of falsely discovered variables. Both theoretical results and  simulations have shown  the correctness of our procedure and improved performance over the Lasso regression. 

For the yeast genotype and gene expression data, our two-stage regression method was able to identify three yeast genes whose expressions were associated growth in the presence 
of  hydrogen peroxide. In contrast, using gene expression data alone and Lasso regression did not identify any growth associated genes.  Since growth yield is highly inheritable  \citep{yeastdrug}, using genotype-predicted gene expressions in our two-stage estimation can help to identify the gene expressions  that might be causal to the phenotype.   For model organisms such as yeast,  the conditional independence assumption between the genotypes and the outcome given  gene expression levels is  expected to hold.  However, for human studies, one should be cautious of such an assumption since genetic variants can affect phenotype via other mechanisms such as changing protein structures. 

One possible application of the proposed two-stage regression  is to identify gene expressions that cause diseases by jointly analysis genotype and gene expression data.  This is similar in spirit to PredXscan \citep{Xscan}  that aims to  identify the molecular mechanisms through which genetic variation affects phenotype. PredXscan builds gene expression prediction models using reference eQTL data. In contrast, our method requires that the genotype and gene expression data are measured on the same set of individuals.

Potential extensions of this paper include detecting and accounting for the existence of weak instrumental variables and developing methods that are robust to the residual distributions. Recent papers such as \citet{chatterjee2010asymptotic} and \citet{dezeure2017high} developed   bootstrapping inference methods for Lasso estimator. It is possible to apply such ideas to the high dimensional IV model considered in this paper.  Besides the two-stage least square method we developed here, an alternative to estimating the parameters in IV model is  by estimating equations. The two-stage least square methods provides optimal estimator under proper model assumptions while the estimating equation is expected to be robust. The problem of testing a single parameter  using  estimating equation under high-dimensional setting has been explored by \citet{neykov2018unified}. It is interesting  to consider the multiple testing procedure when estimating equations are used for estimating the parameters in high-dimensional IV models. 

\section{Supplemental Materials}
The Supplemental Materials include proofs of lemma \ref{lemma: theta}, theorem \ref{thm: T_i}, \ref{thm: FDR}, \ref{thm: FDV}. The Matlab codes used to implement the algorithm and the real data sets will be provided upon request.

\section*{Acknowledgments}
This research was supported by NIH grant GM129781.

\setstretch{1.24}
\bibliographystyle{jasa}
\bibliography{ivfdrref}

\end{document}